\documentclass[a4paper,fleqn,usenatbib,useAMS]{mnras}

\usepackage{graphicx, natbib,epsfig,amsmath,amssymb, mathrsfs, txfonts}
\usepackage{tikz}
\usetikzlibrary{shapes,arrows}

\usepackage[T1]{fontenc}
\usepackage{ae,aecompl}

\usepackage{newtxtext,newtxmath}

\title[High-contrast imaging with deformable mirrors]{\textit{High-contrast imaging at small separation: impact of the optical configuration of two deformable mirrors on dark holes.}}

\author[M. Beaulieu et al]{{M. Beaulieu$^{1}$\thanks{Contact e-mail: \href{mailto:mathilde.beaulieu@oca.eu}{mathilde.beaulieu@oca.eu}},  L. Abe$^{1}$,  P. Martinez$^{1}$,  P. Baudoz$^{2}$,   C. Gouvret$^{1}$, F. Vakili$^{1}$}
\\
$^{1}$Universit\'e C\^{o}te d'Azur, OCA, CNRS, Laboratoire Lagrange, Parc Valrose, F-06108, Nice, France \\
$^{2}$LESIA, Observatoire de Paris, PSL Research University, CNRS, Sorbonne Universit\'{e}s, UPMC Univ. Paris 06, Univ. Paris Diderot, Sorbonne Paris Cit\'{e}, France}

\date{Last updated 2015 May 22; in original form 2013 September 5}

\pubyear{2016}

\begin{document}
\label{firstpage}
\pagerange{\pageref{firstpage}--\pageref{lastpage}}
\maketitle
 
\begin{abstract}
The direct detection and characterization of exoplanets will be a major scientific driver over the next decade, involving the development of very large telescopes and requires high-contrast imaging close to the optical axis. Some complex techniques have been developed to improve the performance at small separations (coronagraphy, wavefront shaping, etc). In this paper, we study some of the fundamental limitations of high contrast at the instrument design level, for cases that use a combination of a coronagraph and two deformable mirrors for wavefront shaping. In particular, we focus on small-separation point-source imaging (around \mbox{1 $\lambda$/D}). First, we analytically or semi-analytically analysing the impact of several instrument design parameters: actuator number, deformable mirror locations and optic aberrations (level and frequency distribution). Second, we develop in-depth Monte Carlo simulation to compare the performance of dark hole correction using a generic test-bed model to test the Fresnel propagation of multiple randomly generated optics static phase errors. We demonstrate that imaging at small separations requires large setup and small dark hole size. The performance is sensitive to the optic aberration amount and spatial frequencies distribution but shows a weak dependence on actuator number or setup architecture when the dark hole is sufficiently small (from 1 to $\lesssim$ \mbox{5 $\lambda$/D}).
\end{abstract}
%



\begingroup
\let\clearpage\relax
\endgroup
\newpage

\section{Introduction}
The direct detection and characterization of exoplanets will be a major scientific driver over the next decade, especially regarding the development of extremely large telescopes (ELTs). High-contrast imaging provides an ideal method to characterize extra-solar planetary systems \citep{MaroisMacintoshBarmanEtAl2008, KalasGrahamChiangEtAl2008, LagrangeBonnefoyChauvinEtAl2010} but faces multiple challenges. In particular, reaching small angular separation between planets and stars is a must if one wants: (1) to focus on exoplanet detection ultimately down to terrestrial planets and very young giant planets, or more pragmatically (2) to take full advantage of the angular resolution of the telescope The simulated planet population for young and nearby star samples \citep{BonavitaChauvinDesideraEtAl2012} shows that an imaging contrast of $10^{-8}$ ($J$ band) must be achieved at a separation of tens of milliarcseconds in order to detect rocky planets (with the ELTs). In this context, the abundance of M dwarfs in the Milky way and their large fraction of low-mass companion \citep{CassanKubasBeaulieuEtAl2012} make them good candidates for searching for young planets, although they show unfavourable properties. \\
\indent 
Small-angular-separation imaging requires a major technological breakthrough because it needs very high image quality and stability, as a large amount of the on-axis point spread function (PSF) is concentrated inside \mbox{1 $\lambda$/D}. High-contrast imaging needs multiple step corrections, where seeing-limited PSFs will constantly be improved by extreme-adaptive optics (ExAO), non-common path aberration control, diffraction suppression or coronagraphy, and science image post-processing to correct for atmospheric, static and quasi-static aberrations. A few coronagraphs reach high-contrast levels at small separation (vortex coronagraph -- \citealt{MawetRiaudAbsilEtAl2005, FooPalaciosSwartzlander2005}, phase-induced amplitude apodizaton -- \citealt{GuyonMartinacheBelikovEtAl2010}) at the cost of high sensitivity to aberrations. While atmospheric aberrations are corrected with an adaptive optic system, various techniques such as PSF subtraction or wavefront control and shaping have been developed to minimize the static or quasi-static part of the aberrations. 
Quasi-static speckle calibration through post-processing or observational strategies is routinely exploited in all leading observatories equipped with exoplanet hunter instruments. For instance, GPI \citep{MacintoshGrahamPalmerEtAl2007}, SPHERE \citep{BeuzitFeldtDohlenEtAl2008} and SCExAO \citep{GuyonMartinacheGarrelEtAl2010} make use of the most recent and well-known techniques such as spectral differential imaging (SDI, \citealt{MaroisLafreniereDoyonEtAl2006}), spectral deconvolution (SD, \citealt{SparksFord2002}), angular differential imaging (ADI, \citealt{MaroisDoyonRacineEtAl2005}), or polarimetric differential imaging (PDI, \citealt{KuhnPotterParise2001}). These techniques significantly increase the sensitivity of observing sequences, but become inefficient for small-angle field of view. At small inner working angles (IWA), these above-mentioned techniques suffer from insufficient chromatic speckle elongation (with SD, speckles must move by at least one resolution element between the shortest and the longest wavelength, and thus it tends toward a large spectral range), and an insufficient field of rotation (with ADI, speckles must rotate by at least one resolution element, which translates into long observing times, for which temporal decorrelation of quasi-static speckles will occur as a fundamental limitation). In addition, SDI relies on an unknown a priori feature in the exoplanet's spectrum and is sensitive to non-common path errors and PDI, despite a higher sensitivity at small separations (owing to increased reflected light), is severely impacted by photon noise because unpolarized reflected light is not used. Small IWA observing mode thus call for alternative approaches or observing strategies, especially in relation to ELTs. \\
\indent 
Wavefront shaping is an alternative way to address this challenging issue. Rather than the wavefront control being the process of flattening wavefront errors from imperfect optics, we refer, by wavefront shaping, to the process of creating a dark zone (the so-called dark hole) in the PSF. In this context, the calibration of quasi-static speckles at small IWA can be done at a reduced efficiency through coherence-based methods that can be implemented by modulating the light in the speckles, either temporally \citep{Guyon2004} or spatially \citep{BaudozMazoyerMasEtAl2012}. One limitation of wavefront shaping is the Fresnel propagation of phase aberrations, described by the `Talbot effect': at the deformable mirror (DM) plane, out-of-pupil optics create a mix of amplitude and phase errors that a single DM cannot correct on the full field, or only at the expense of loosing at least half of the field (e.g. \citealt{BordeTraub2006, GiveonKernShaklanEtAl2007}. One way to tackle this effect is to use at least two DMs to correct for both phase and amplitude. Multiple-DM control has been under extensive testing worldwide for more than a decade in many laboratory test-beds (THD, \citealt{GalicherBaudozDelormeEtAl2014}; HCIT, \citealt{RiggsGroffCarlottiEtAl2013}; HCIL, \citealt{PueyoKayKasdinEtAl2009,RiggsGroffCarlottiEtAl2013}; HiCAT, \citealt{NDiayeChoquetPueyoEtAl2013}; and SPEED, \citealt{MartinezPreisGouvretEtAl2014}) that come within the scope of future on-sky applications. A set of deformable mirrors is used to correct for the wavefront error from imperfect optical surfaces as well as to shape the wavefront to produce a dark zone in the PSF halo.
Various successful laboratory experiments using either a single DM (e.g \citealt{TraugerTraub2007, GuyonPluzhnikMartinacheEtAl2010, BelikovPluzhnikConnelleyEtAl2010,MazoyerBaudozGalicherEtAl2014,DelormeNDiayeGalicherEtAl2016}) or two DMs (e.g., \citealt{KayPueyoKasdin2009, PueyoKayKasdinEtAl2009, RiggsGroffCarlottiEtAl2013}) are appealing for on-sky implementation of the technique. Nonetheless, a well-developed understanding and mastery of multiple-DM architecture is to our knowledge limited to large and/or moderate IWA science goals, leaving the slot of very small IWA uncovered. It is worth exploring wavefront shaping optimization for this specific scientific window, which will ultimately allow the successful implementation of the small IWA mode.\\
\indent
In this paper, we present an in-depth understanding of science-grounded instrument conception and contrast design at small IWA (\mbox{1 $\lambda$/D}). The relationship between scientific and instrumental requirements is not trivial, especially considering the Fresnel/Talbot perturbations, and is generally not addressed when the optical design is defined. The scope of this work is essentially to assess the relative impact of several optical setup parameters on the ability of two DMs to efficiently control phase and amplitude to create dark holes, as the optics' Fresnel propagation constrains the use of multiple DMs. A general background detailing current dark hole algorithms and architecture with two DMs is presented in \mbox{Section \ref{sec:gen}}. The implemented parameters, which potentially drive the Fresnel propagation effects and contribute to limiting the dark hole depth are related to the DM configuration (number of actuators, location, configuration, etc.) and the optical component quality (wavefront error amount, frequency distribution). Specific limitations  (DM location, number of actuators and aliased speckles) are illustrated and explained with simple cases in \mbox{Section \ref{sec:limitation}}, and \mbox{Section \ref{sec:results}} gathers Fresnel-propagation simulation outputs resulting from a Monte Carlo approach for each of these parameters or combination of parameters. In the scientific context of detecting exoplanets in their habitable zone with future ELTs from the ground, or with dedicated missions in space, high image quality (with ExAO on ground) and high-contrast coronagraph performance are expected. We thus consider, for this study, some generic \textit{perfect coronagraph} (\mbox{Section \ref{sec:model}}), meaning that we focus on more \textit{intrinsic} properties of the optics and setup (polishing frequency distribution, relative beam size, distance between optics, and especially between DMs relative to the pupil plane). In the same way, we assume a perfect AO system (sensing and correction) to emphasize the impact of optical setup parameters. The contrast obtained in Sections \ref{sec:limitation} and \ref{sec:results} ($\thicksim10^{-14}$ at best) is thus well below what real instruments can achieve; for instance, the photon noise from AO residual aberrations is already expected to prevent high-contrast instruments from reaching below $10^{-9}$ (even for the next very large-aperture ELTs, \citealt{Kasper2012}). Nevertheless we show that, if some of the above-mentionned optical parameters are not appropriately set (e.g the deformable mirror location), this can limit the contrast level to $\thicksim10^{-7}$.

\section{General background}
\label{sec:gen}
This section introduces existing model-dependent techniques, namely algorithm implementation and optical architecture, to create a local dark hole at the image plane. 

\subsection{Dark hole algorithm}
We focus on dark hole algorithms when assuming a linear response to optical system perturbations. Wavefront shaping algorithms that deal with non-linearity, for example to correct for pupil discontinuities \citep{PueyoNorman2013}, are out of the scope of this paper.  We describe the existing algorithms (speckle nulling and energy minimization), and the analytical equation for energy minimization (with one and two DM(s)).

\subsubsection{Iterative speckle nulling}
The classical speckle nulling technique \citep{TraugerBurrowsGordonEtAl2004} has proved its performance for laboratory setup (HCIT) and on-sky instruments (GPI, SCExAO). This method is based on identifying the brightest peckles at the focal plane and determining their corresponding phase at pupil plane. The phase/speckle relation is recovered by applying different phase shapes (sinusoidal shape) to the DM and tracking speckle intensity variations at the focal plane. The brightest speckles are thus iteratively removed. The drawback of this method is that it can only remove speckles at the focal plane with frequencies inside the dark hole, and thus it cannot correct for those aliased speckles outside the correction range that create features of sizes smaller than the PSF core (for details see Section \ref{sec:folding}).
 
\subsubsection{Energy minimization}
\label{sec:energy}
Another method redefines the problem by minimizing the energy inside the dark hole. The analytical approach of this method is described  in \citet{GiveonKernShaklanEtAl2007}, \citet{PueyoKayKasdinEtAl2009} and \citet{Groff2012} and is defined by first computing the total energy at the image plane with one DM and then generalizing to the case with two DMs. \\
We assume here a single DM correction (at the pupil plane), an entrance aperture \textit{A}, an initial aberrated field \textit{$\varphi$} (\textit{$\varphi$} can represent both phase and amplitude errors, but for simplicity we assume only phase error) and the DM perturbation $\psi_{\mathrm{DM}}$ such that the electric field at the pupil plane is given by
\begin{equation}
\mathrm{
E_{p}(u,v)= A(u,v) e^{i\varphi} e^{i\psi_{DM}(u,v)},
}
\end{equation}
where \textit{u} and \textit{v} are the spatial coordinates at the pupil plane.
We define \textit{C} as the linear operator from the pupil plane to the image plane such that the final electric field is given by $\mathrm{E_{f}(x,y)= C\{E_{p}(u,v)\}}$ where \textit{x} and \textit{y} are the spatial coordinates at image plane. By assuming small phase and DM perturbation (linear approximation \mbox{${e^{i\varphi} \sim 1 + i\varphi}$}) and by dropping the second-order terms, the focal plane amplitude can be written as
\begin{equation}
\mathrm{
E_{f}(x,y)\thicksim C\{A(u,v) e^{i\varphi}\} + i C\{A(u,v) \psi_{DM}(u,v)\}.}
\label{eq:eqc}
\end{equation}
The intensity inside the dark hole, \textit{$I_{\mathrm{DH}}$}, can be written as \citep{Groff2012}
\begin{align}
\mathrm{I_{DH}(x,y)} & \mathrm{= \iint_{x,y \in DH} E_f(x,y) E_f^*(x,y)\,\mathrm{d}x \mathrm{d}y,} \\
\begin{split}
& \mathrm{= \langle C\{Ae^{i\varphi}\},C\{Ae^{i\varphi}\} \rangle + \langle C\{A\psi_{DM}\},C\{A\psi_{DM}\} \rangle} \\
& \mathrm{+2 \Im(\langle C\{Ae^{i\varphi}\},C\{A\psi_{DM}\} \rangle),} \\
\end{split}
\end{align}
where $\Im$ represents the imaginary part and \textit{*} is the complex conjugate. This equation can be written as a matrix multiplication by assuming that the DM phase is described by the DM actuator number \textit{$N$}, the DM influence functions \textit{$f_k$} and the DM coefficients \textit{$a_k$} such that  \mbox{$\mathrm{\psi_{DM}=\sum_{k=0}^{N} a_{k} f_{k}(u,v)}$}.
The intensity inside the dark hole is thus
\begin{align*}
 \mathrm{I_{DH}= {^t}a \ M_0 \ a + 2 \ {^t}a  \ \Im(b_0) + d_0,} \\
 \mathrm{where} \quad & \mathrm{M_0=\langle C\{Af\},\ C\{Af\} \rangle = G^*G,} \\ 
 &  \mathrm{G=C\{Af\},} \\
 & \mathrm{b_0=G^*C\{Ae^{i\varphi}\},} \\
 & \mathrm{d_0=\vert C\{Ae^{i\varphi}\} \vert^2.}
\end{align*}
$M_0$ represents the system response to each DM poke, $b_0$ represents the interaction between the DM and the aberration, and $d_0$ is the initial intensity owing to aberrations.
The algorithm minimizes the intensity inside the dark hole by nulling its derivative with 
\begin{equation}
\mathrm{
\mbox{$a=-M_{0}^{-1}\Im(b_0)$.}}
\label{eq:coeff}
\end{equation}
Resolving equation \eqref{eq:coeff} can lead to solution with large stroke values. One way to limit the algorithm to stable solutions is to use the electric field conjugation (EFC) or stroke minimization method.
EFC \citep{GiveonKernShaklanEtAl2007} minimizes the dark hole intensity \textit{$I_{\mathrm{DH}}$} weighted by a Tikhonov regularization. The cost function to minimize is
\begin{equation}
\mathrm{
\mbox{$J=I_{DH}+\alpha_0^2 {\Vert a \Vert}^2$},}
\end{equation}
where $\alpha_0$ is the Tikhonov regularization parameter that guarantees that the algorithm converges within stable actuator stroke values. The DM coefficients are defined as
\begin{equation}
\mathrm{
\mbox{$a=-(M_{0}+\alpha_0^2 1\!\!1)^{-1}\Im(b_0)$,}}
\end{equation}
where \mbox{$1\!\!1$} is the identity matrix. The parameter $\alpha_0$ represents the actuator stroke weight in the minimization and is defined by linearly increasing $\alpha_0$ and finding the smallest value that achieves high contrast. EFC thus seeks to find the minimum energy within some weighted stroke solution. \\
A second approach, stroke minimization \citep{PueyoKayKasdinEtAl2009}, seeks to find the minimum stroke values that reach a given contrast ratio. It minimizes the DM coefficients \mbox{${\Vert a \Vert}^2$} such that the intensity in the dark hole \mbox{$I_{\mathrm{DH}} \leqslant 10^{-\mathbb{C}}$} where $\mathbb{C}$ is the targeted contrast ratio. The cost function to minimize is
\begin{equation}
\mathrm{
\mbox{$J={\Vert a \Vert}^2 + \mu_0 (I_{DH} - {10}^{-\mathbb{C}})$.}}
\end{equation}
The DM coefficients are computed using
\begin{equation}
\mathrm{
\mbox{$a=-(M_{0}+\frac{1}{\mu_0}1\!\!1)^{-1}\Im(b_0)$.}}
\end{equation}
As for the EFC, the parameter $\mu_0$ is determined by line search on  $\mu_0$. The EFC and stroke-minimization methods are equivalent for a single DM in monochromatic light. \\

\cite{PueyoKayKasdinEtAl2009} and \cite{Groff2012} demonstrated that the case with one DM can be generalized to two DMs by defining the linear operators $C_{1}$ and  $C_{2}$ from respectively $\mathrm{DM_1}$ (first DM) and $\mathrm{DM_2}$ (second DM) to the image plane\footnote{In the following, when referring to DM distances, $\mathrm{DM_1}$ is always located upstream of the pupil plane while $\mathrm{DM_2}$ is always downstream.}. The interaction matrix becomes $M_0=G^*G$ with $G=[ G_1, G_2]$.

The energy minimization method reformulates the problem in a global way by correcting for the overall energy inside the dark zone and thus partially address the issue of aliased speckles. A speckle at a frequency outside the DM correction range cannot be fully extinguished as the DM cannot mimic its central frequency, but its intensity can be decreased by fitting a linear combination of speckles inside the correction range (see Section \ref{sec:folding}).
 
\subsection{Setup architecture}
\label{archi}
\begin{figure*}
\centering
\includegraphics[height=4 cm]{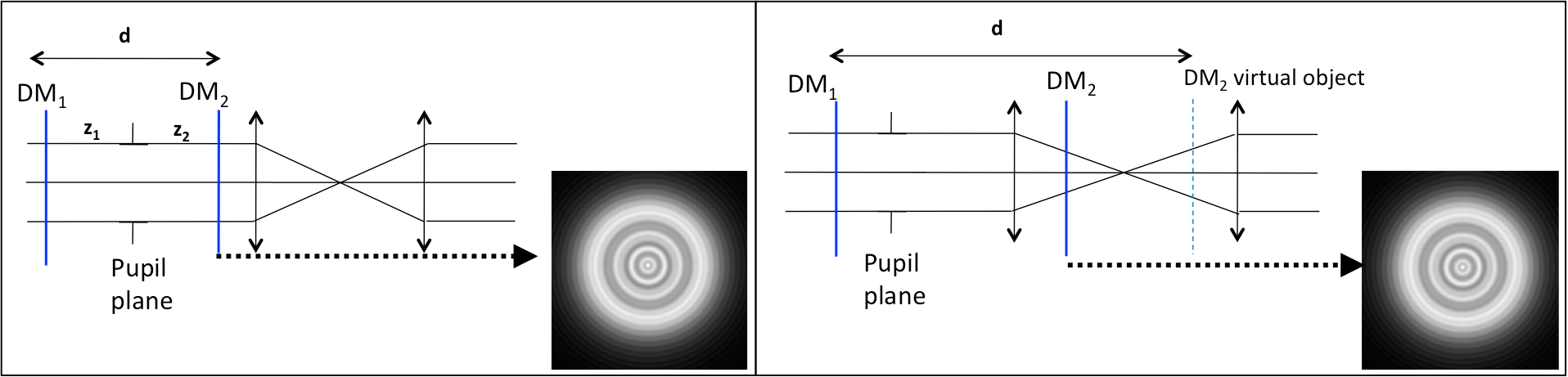}
\caption{Setup with one deformable mirror in a collimated (left) and converging (right) beam, and their corresponding amplitude pattern.}
\label{fig:archi0}
\end{figure*}
Wavefront shaping with two DMs can be implemented in various ways: in collimated or convergent beams. \citet{ZhouBurge2010} showed that diffraction propagation effects can be computed in a converging beam as well, using the conjugation location of elements in an equivalent unfolded collimated beam design. The two setups are described in Fig.~\ref{fig:archi0}. The case of the converging beam (right in the figure) is defined with $\mathrm{DM_2}$ located after the lens but before the focus, such that the DM virtual object is after the focal length, leading to a large equivalent distance between the two DMs (\textit{d}). The converging beam setup parameters are determined using geometric optics by defining the geometrical beam size at $\mathrm{DM_2}$. The $\mathrm{DM_2}$ amplitude patterns shown in Fig.~\ref{fig:archi0} are computed with Fresnel propagation (\textsc{proper} code, \citealt{Krist2007a}) for the two architectures with equal equivalent distance between the two DMs, showing the same diffracting structure. The main advantage of the converging beam setup is a large gain in total setup length, leading, however, to a potential loss in the wavefront shaping efficiency, as the beam size at $\mathrm{DM_2}$ is smaller than the case for the collimated beam and thus contains less DM actuators. For instance, in the generic setup used for the main simulation (see Section \ref{opt_mod}), a loss of 10\% actuators leads to a more compact optical train of about 15\%. The converging beam setup can be used when the physical size of one DM is smaller than the other one and thus cannot be implemented in the collimated beam, but the performance of the two architectures has not yet been compared.

\section{Potential fundamental limitation to high contrast at small separation}
\label{sec:limitation}
The previous section presented the general background for creating a dark hole using two DMs (algorithm and setup architecture). The following section describes and illustrates, using simple cases, the setup limitations for high contrast at small separations (DM location in the setup, DM actuator number and aliased speckles). 
\subsection{Setup DM location}
\label{dmloc}
This section describes how to determine the optimal DM distances for high-contrast imaging at small separation (around \mbox{1 $\lambda$/D}). For that purpose, we analytically compute the impact of an out-of-pupil DM in a simple imaging setup.
\subsubsection{Impact of one out-of-pupil DM}
We assume a simple imaging setup with one lens (focal length \textit{F}) such that the linear operator from the pupil plane to the image plane (\textit{C} in Section \ref{sec:energy}) is a Fourier transform. A DM is located at a distance \textit{z} downstream from the pupil plane.
The electric field at the focal plane ($\mathrm{E_f}$) is computed by first Fresnel propagating the DM electric field to the lens denoted $\mathrm{E_l(\alpha,\beta)}$ where $\alpha$ and $\beta$ are the spatial coordinates, adding the lens contribution and finally propagating to the focal plane. The electric field at the lens plane is defined as
\begin{equation}
\mathrm{
\label{eqdm}
E_{l}(\alpha,\beta)=\frac{e^{i\frac{2\pi}{\lambda}(F-z)}}{i\lambda(F-z)}\iint E_{DM}(u,v)e^{i\frac{\pi}{\lambda(F-z)}((\alpha-u)^2+(\beta-v)^2)}\mathrm{d}u\mathrm{d}v, }
\end{equation}
where \textit{$\mathrm{E_{\tiny{DM}}}$} is the DM electric field (DM amplitude and phase perturbation depending on the spatial coordinates \textit{u} and \textit{v}). The complex amplitude $\mathrm{E_{f}(x,y)}$ at the focal plane is thus
\begin{equation}
\mathrm{
E_f(x,y)=\frac{e^{i\frac{2 \pi F}{\lambda}}}{i\lambda F}e^{i\frac{\pi z}{\lambda F}(x^2+y^2)}} \  \widehat{\mathrm{E_l(\alpha,\beta)},}
\end{equation}
\begin{multline}
\mathrm{E_f(x,y)  = \frac{e^{i\frac{2\pi}{\lambda}(2F-z)}}{i \lambda F}  \ [cos(\frac{\pi z}{\lambda F^2}(x^2+y^2))} \\ 
 \mathrm{\quad + i \ sin(\frac{\pi z}{\lambda F^2}(x^2+y^2))]} \ \widehat{\mathrm{E_{\scriptscriptstyle{DM}(u,v)}}},
 \label{eq:dm01}
\end{multline}
where \ $\widehat{}$ \  represents the Fourier transform. The electric field, when expressing the spatial frequencies in units of \mbox{$\lambda$ / D} (where \textit{D} is the pupil diameter), is given by
\begin{multline}
\mathrm{
E_f(x',y')  = \frac{e^{i\frac{2\pi}{\lambda}(2F-z)}}{i \lambda F}   \ [cos(\frac{\pi \lambda z}{D^2}(x'^2+y'^2))} \\ 
 \mathrm{\quad + i \ sin(\frac{\pi \lambda z}{D^2}(x'^2+y'^2))]}  \ \widehat{\mathrm{E_{\scriptscriptstyle{DM}(u,v)}}},
 \label{eq:dm1}
\end{multline}
where \textit{x'} and \textit{y'} are the spatial frequencies in units of \mbox{$\lambda$ / D}.
The image-plane electric field is thus modulated by sine and cosine functions depending on the DM location \textit{z} but also on the dark hole spatial frequencies (image-plane coordinates \textit{x} and \textit{y}).
\begin{figure}
\centering
\includegraphics[width=\columnwidth]{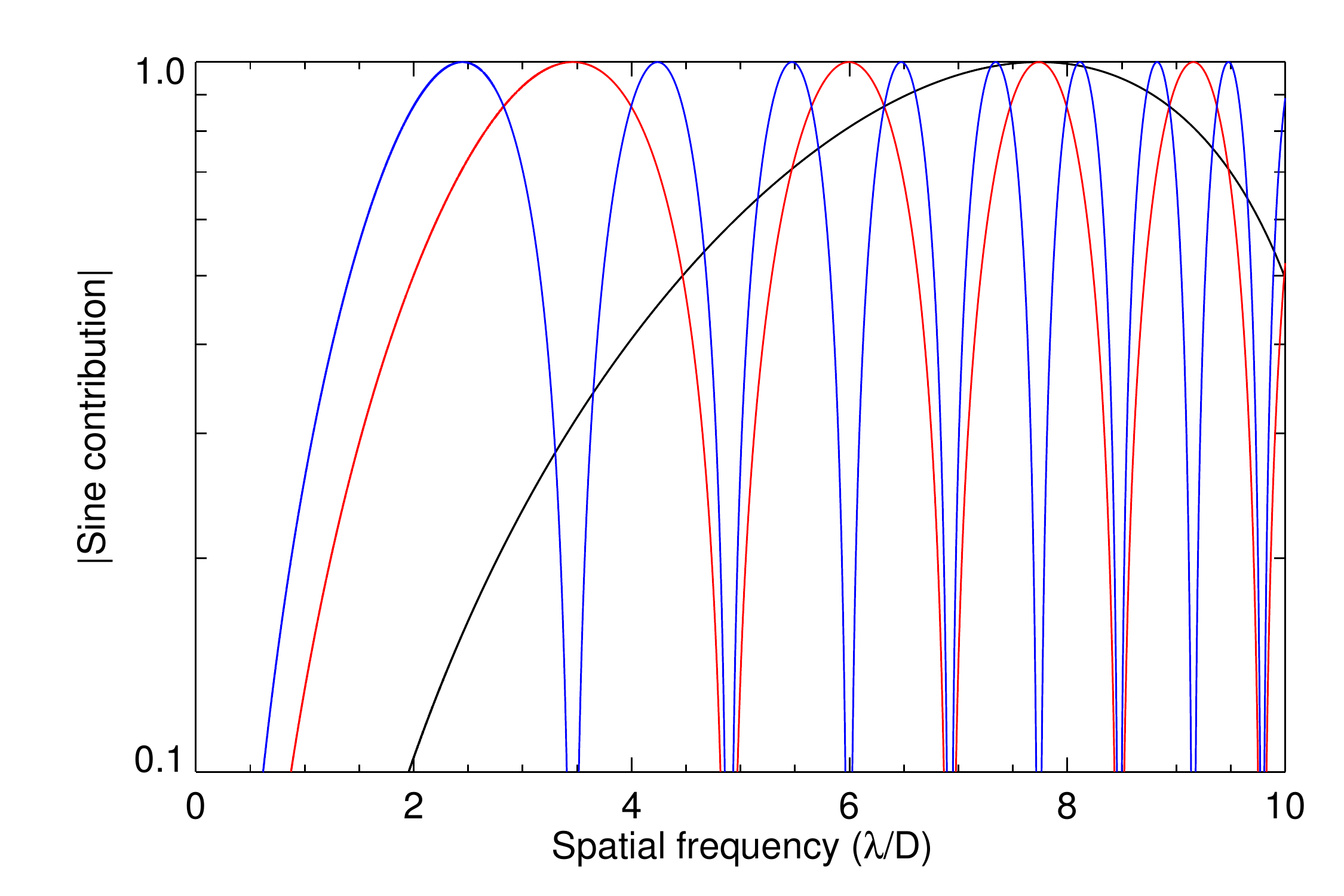}
\caption{Sine contribution as function of the dark hole frequencies for different deformable mirror locations: 0.3 m (black), 1.5 m (red) and 3 m (blue) from the pupil plane.}
\label{fig:dmlocation0}
\end{figure}
\begin{figure}
\centering
\includegraphics[width=\columnwidth]{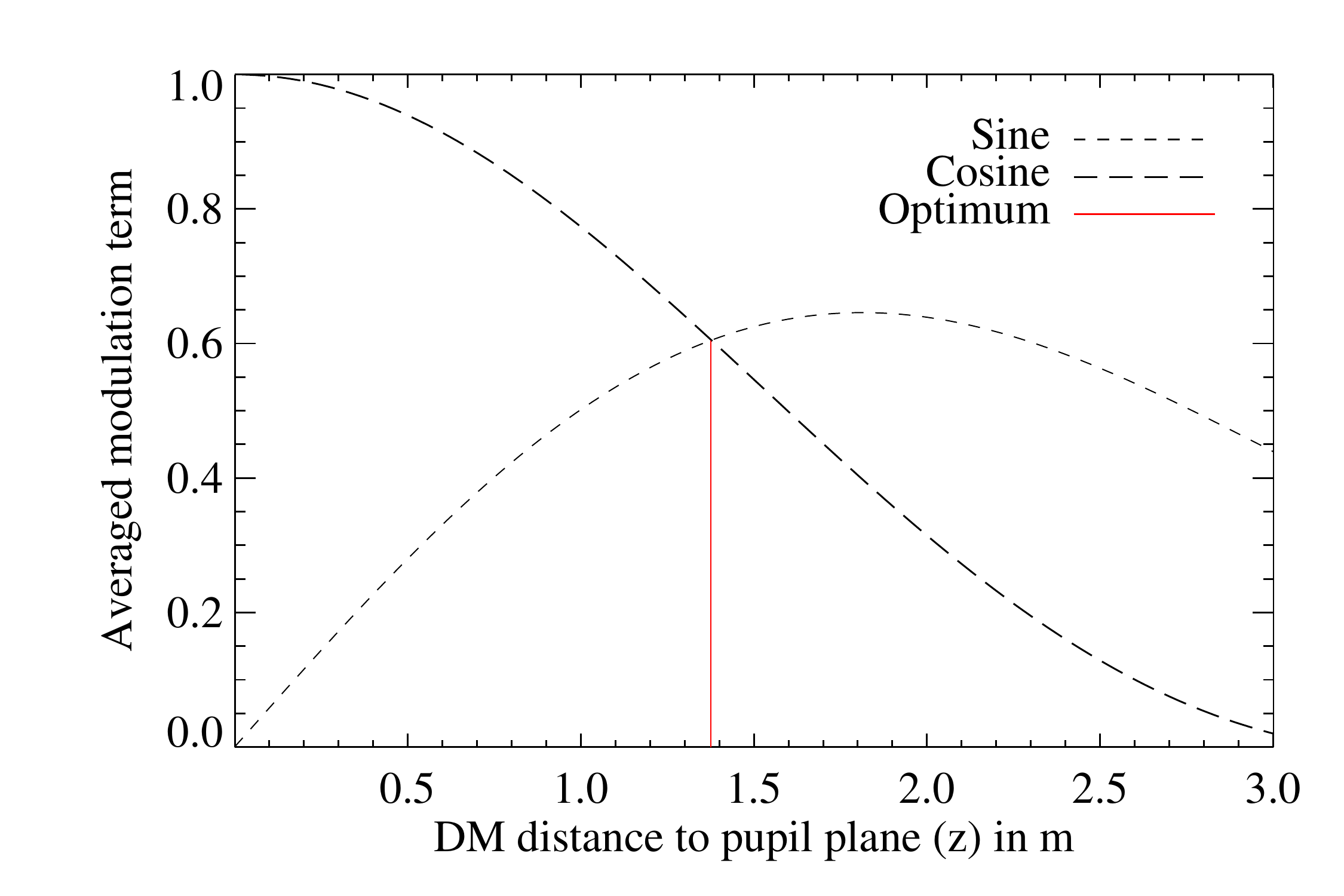}
\caption{Sine (dash) and cosine (long dash) contributions averaged from 0.8 to 4 $\lambda$/D as a function of the deformable mirror location \textit{z} in metres and optimum deformable mirror location (red line) that maximizes the overall coverage.}
\label{fig:dmlocation01}
\end{figure}
This modulation impacts the real and imaginary parts of the image plane and contributes to the DM efficiency in the sense of stroke amplitude: low sine or cosine values will need to be compensated by a large DM stroke, which will be outside of the algorithm linear regime assumption described in Section \ref{sec:energy}, and outside of the DM correction range. We first focus on the sine term by showing (see Fig.~\ref{fig:dmlocation0}) the absolute values of the sine contribution for three DM locations (0.3 m (black), 1.5 m (red) and 3 m (blue)), in the case of a beam diameter of \mbox{7.7 mm} and a wavelength of \mbox{1.65 $\mathrm{\mu m}$} (for consistency with the more complex model described in \mbox{Section \ref{sec:model}}). It can be seen that the sine term oscillates and thus degrades the overall efficiency over the spatial frequencies. We also note that the DM location can be optimized to maximize the averaged sine coverage over a defined dark hole. In particular, the sine contribution for high-contrast imaging at large separation (around 4 to \mbox{$10 \lambda/D$}) is well covered by a DM located at \mbox{0.3 m} from the pupil plane (black curve). On the other hand, high-contrast imaging at small separation (around \mbox{$1 \lambda/D$}) requires large DM location (\mbox{$>$1 m}, blue and red curves) but also a small dark hole size: the sine contribution for those DM locations shows large oscillations and thus poor coverage when trying to enlarge the dark zone region. We can apply the same rationale for the cosine contribution: Fig.~\ref{fig:dmlocation01} shows the modulation contribution (sine and cosine) averaged over spatial frequencies from 0.8 to \mbox{4 $\lambda$/D} (for a goal of high contrast imaging at small separation). The optimum distance is the one with the same contribution for the real and imaginary parts of the focal plane (the intersection of the sine and cosine contribution, shown in red in Fig.~\ref{fig:dmlocation01}). The modulation term thus directly impacts the DM stroke, even more for a high initial aberration level, and depends on the wavelength and pupil diameter (proportional to the Fresnel number, see equation \ref{eq:dm1}). The discussion part of this paper describes the impact of the pupil diameter and wavelength on the optimum DM distances (see Fig.~\ref{fig:discuss}, Section \ref{sec:discussion}).

\subsubsection{Impact of two out-of-pupil DMs} 
\begin{figure}
\centering
\includegraphics[width=\columnwidth]{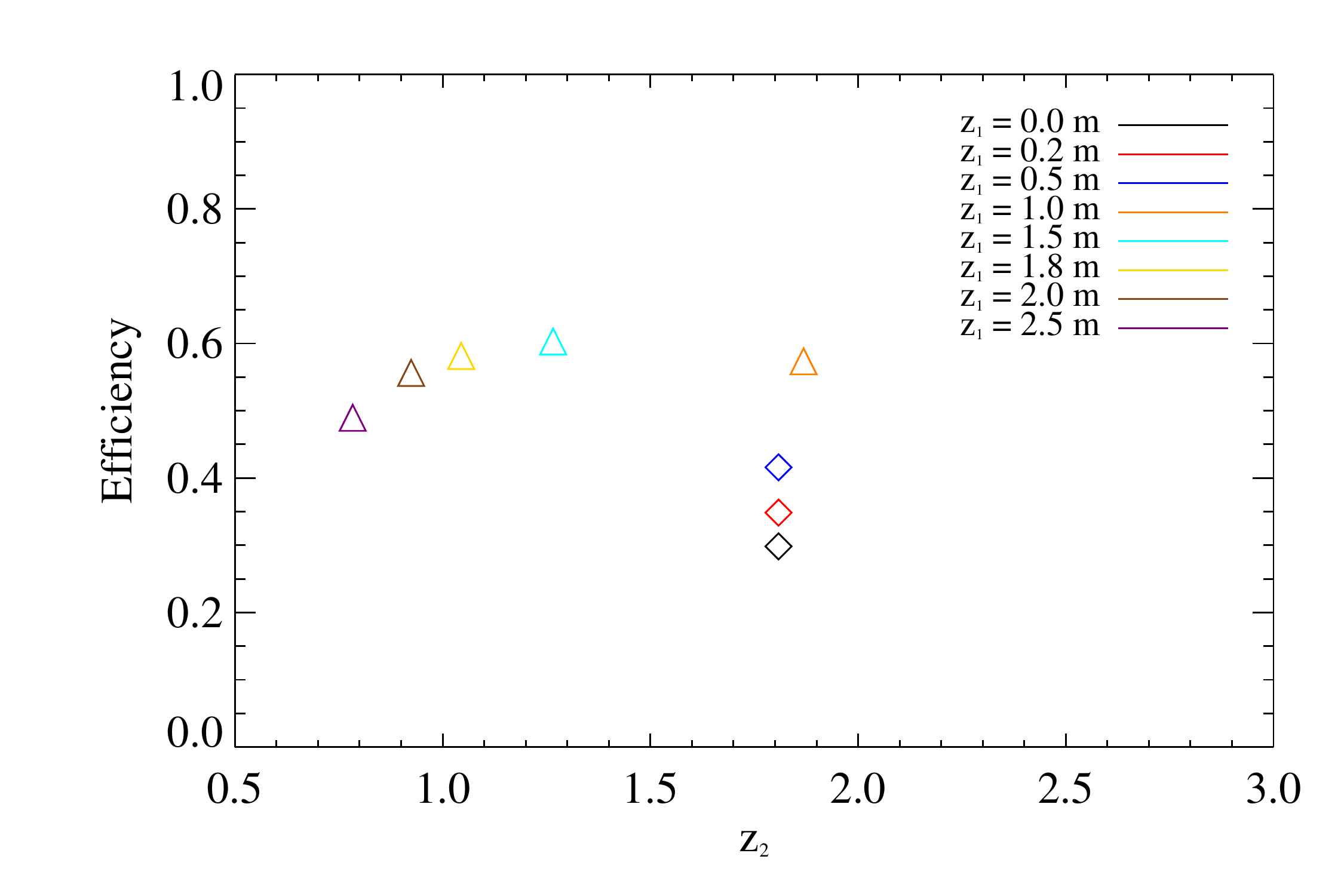}
\caption{Efficiency within 0.8 to 4 $\lambda/D$ as a function of $\mathrm{DM_2}$ location for several $\mathrm{DM_1}$ locations, from the pupil plane to 2.5 m. The triangles correspond to the optimum cases where the sine and cosine contributions are equal whereas the diamonds correspond to cases where the cosine and the sine contributions are not identical for any deformable mirror location. In that case, the optimum distance and the efficiency are defined at the maximum sine contribution.}
\label{fig:dmlocation1}
\end{figure}

The previous section shows that the location of an out-of-pupil DM determines the focal plane modulation impact (cosine and sine contribution) at the dark hole frequencies. The case with two DMs can be derived from the case with one DM by assuming that the DM contributions are small and that there is no diffracting element between the two DMs. We can thus add the two modulation contributions and write the focal plane electric field as
\begin{multline}
\mathrm{
{E_{f}}_{\scriptscriptstyle{2DM}}(x',y') \propto a_1 cos(\frac{\pi \lambda z_1}{D^2}(x'^2+y'^2)+a_2 cos(\frac{\pi \lambda z_2}{D^2}(x'^2+y'^2)} \\
\mathrm{+ i \ \left [a_1 sin(\frac{\pi \lambda z_1}{D^2}(x'^2+y'^2)+ a_2 sin(\frac{\pi \lambda z_2}{D^2}(x'^2+y'^2) \right]}, 
\label{eq:dm2}
\end{multline}
where \textit{$z_1$} and \textit{$z_2$} are the distances of each DM to the pupil plane and \textit{$a_1$} and \textit{$a_2$} represent the contribution of each DM via the Fourier transform of their complex amplitude. In the following, we assume that the two contributions are of the same order of magnitude and thus that we can add the cosine and sine contributions to determine the optimum DM distance. \\
In order to determine the optimum DM locations that correct for small frequencies, we (1) compute the absolute value of the overall sine contribution (sum of the two DMs), (2) compute the absolute value of the overall cosine contribution, and (3) find the optimum distance at the intersection of the two contributions.
The results are shown in Fig.~\ref{fig:dmlocation1} which gives the efficiency for different DM locations. Diamonds correspond to the worst case, as the cosine and the sine contributions never intersect. The sine contribution is thus the efficiency criterion as, at small separations, the sine contribution is always smaller than the cosine contribution and thus limits the performance.  A DM at the pupil plane or near the pupil plane is thus not optimum for high-contrast imaging between 0.8 and \mbox{4 $\lambda$/D}. As for the case with one DM, large setups provide better efficiency for the performance at small separations. The optimum DM locations are for $z_1$ and $z_2$ at \mbox{1.5} and  \mbox{$\thicksim$1.3 m}, with similar performance between 1 and \mbox{2 m}. 

\subsection{DM actuator number}
\label{sec:act_number}
\begin{figure}
\centering
\includegraphics[width=.99\columnwidth]{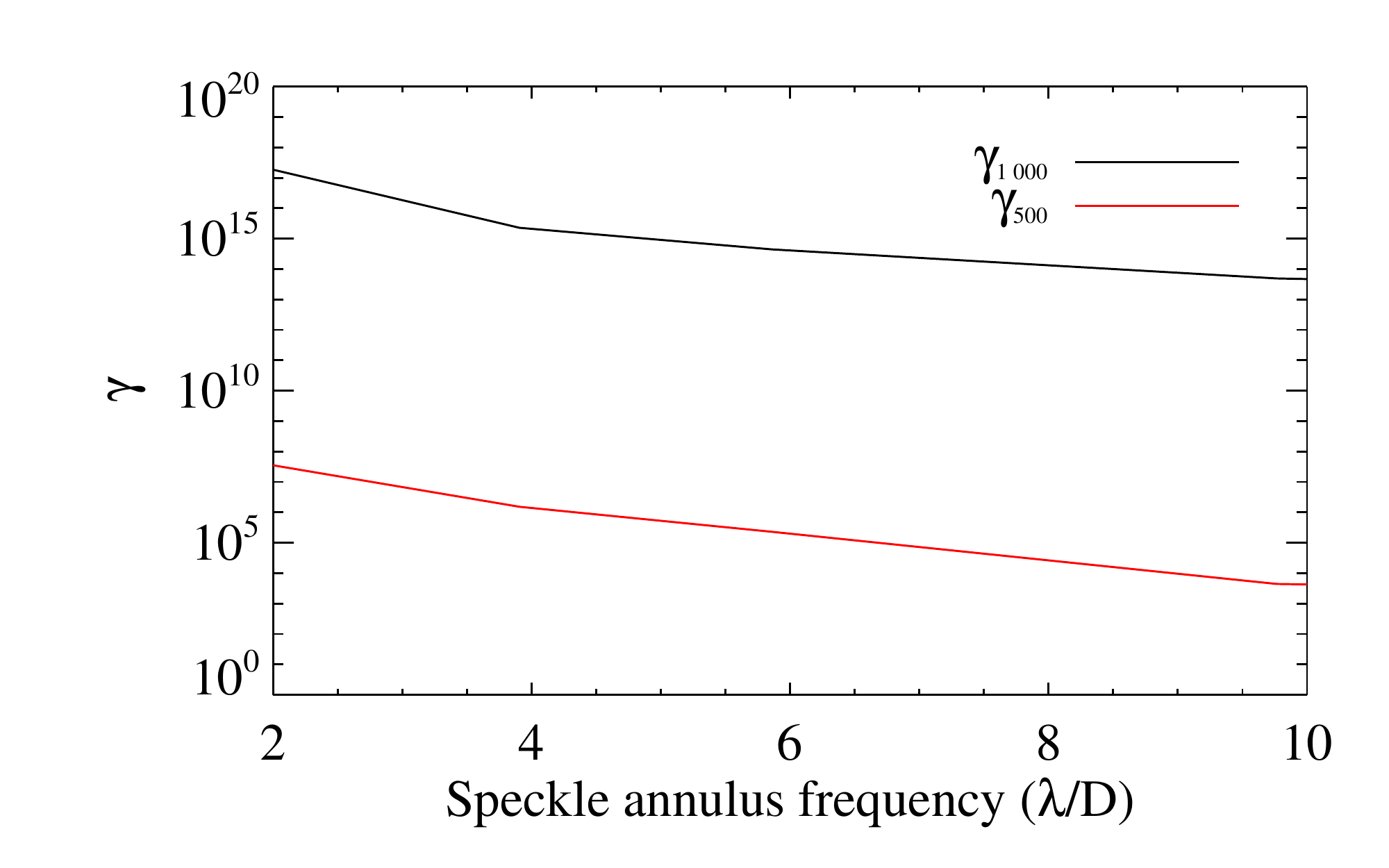}
\caption{Gain in contrast ($\gamma$) when increasing the actuator number from 500 (red) to \mbox{1000} (black), for a dark hole size of \mbox{10 $\lambda$/D}.}
\label{fig:act_number0}
\end{figure}

\begin{figure}
\centering
\includegraphics[width=.99\columnwidth]{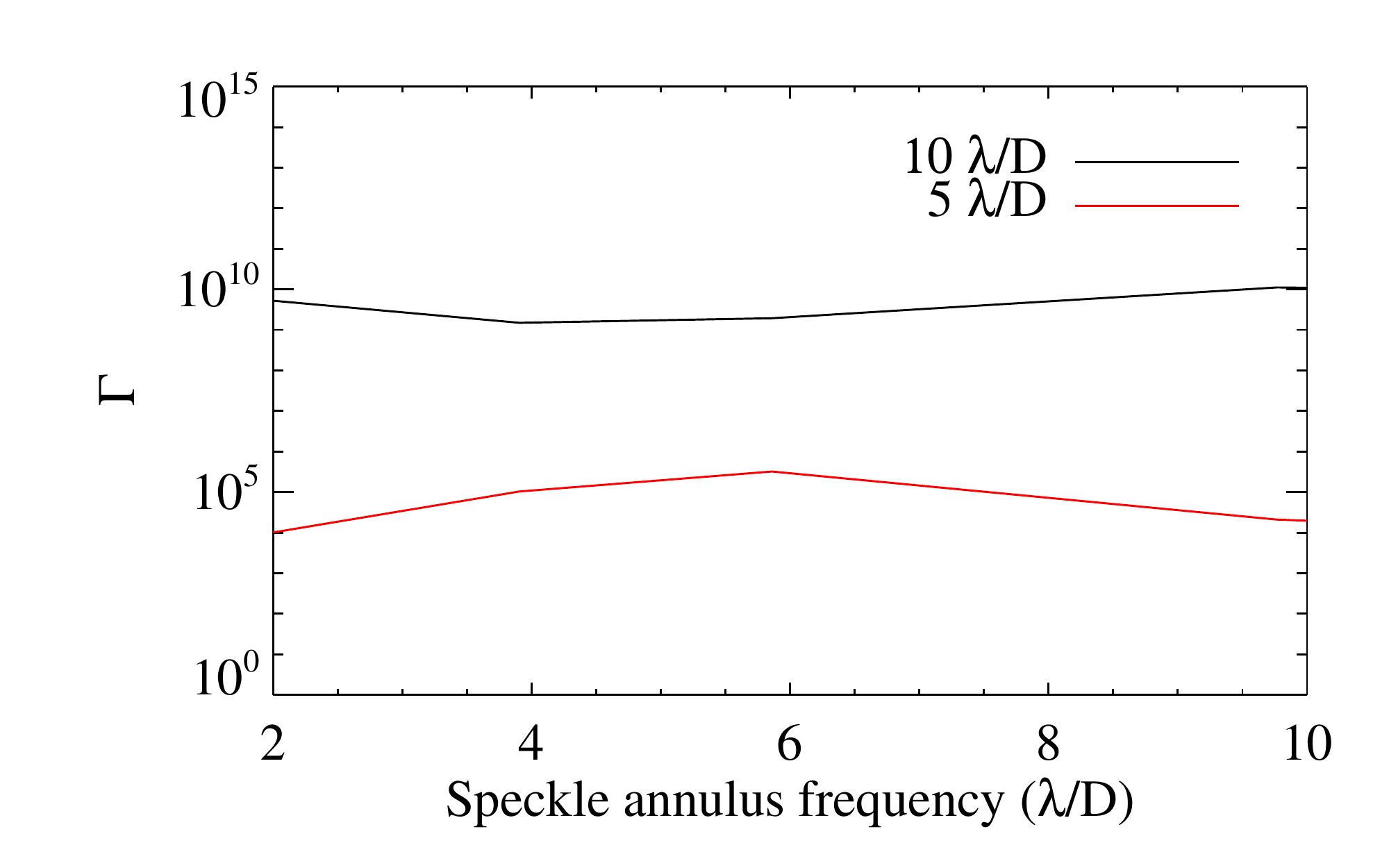}
\caption{Gain in contrast owing to doubling the actuator number ($\Gamma$) for dark hole sizes of \mbox{10 $\lambda$/D} (black) and \mbox{5 $\lambda$/D} (red).} 
\label{fig:act_number1} 
\end{figure}

Because of the Nyquist criterion, the number of actuators \textit{N} limits the DM overall performance: we can correct up to a radius of \mbox{$\lambda N / 2D $} (DM correction range) at the image plane. Furthermore, the number of actuators also impacts high-contrast imaging inside the DM correction range: performance depends on the DM capability to reproduce a phase pattern well, even for spatial frequencies less than \mbox{$\lambda N / 2D $}. To illustrate this point, we simulate a setup with a single DM located at the pupil plane. The relationship between pupil and image planes is a Fourier transform and we assume a perfect coronagraph/setup that removes the real part of the electric field $\mathrm{E_f}$ such that
\begin{equation}
\mathrm{
E_{f}=i}\widehat{\mathrm{A \varphi}.}
\label{eq:fr}
\end{equation}
The initial complex amplitude at the pupil plane is defined by the  phase aberrations \textit{$\varphi(u,v)$} and the pupil shape \textit{A(u,v)}. The phase pattern is a radial cosine (\mbox{cos($\mathrm{2\pi r \nu})$} with \textit{r} as the radial distance and \textit{$\nu$} the frequency), creating an \textit{annulus of speckles} at a given frequency at the image plane (see Section \ref{sec:folding} for an illustration of the annulus of speckle). We assume an aberration amount of 10 nm rms. We can thus assess the system response depending on the defined initial phase frequency. The dark hole algorithm minimizes the energy at the focal plane using equation \ref{eq:coeff} within a dark hole of width \mbox{10 $\lambda$/D}. In order to illustrate the DM capability to correct for aberration frequencies inside the DM correction range, we define the following terms. The contrast ratio $\mathscr{C}$ is the \mbox{$5\sigma$} contrast ratio (detection threshold) computed inside the dark hole of the PSF divided by the maximum value of the initial PSF without a coronagraph or wavefront shaping. We also define what we call the wavefront shaping gain $\gamma$, which is the ratio of the contrast $\mathscr{C}$ before and after the wavefront shaping algorithm. This definition comes from the fact that, as the speckle annulus frequency increases, its contribution inside the dark hole decreases because of the diffraction pattern which is, in this case, the Airy profile) leading to a better contrast ratio not correlated to the wavefront shaping performance. The wavefront shaping gain is thus
\begin{equation}
\gamma = \frac{\mathscr{C}_{\textrm{before}}}{\mathscr{C}_{\textrm{after}}}.
\label{eq:gain}
\end{equation}
 \mbox{Fig.~\ref{fig:act_number0}} illustrates the DM capability to correct for aberration frequencies inside the DM correction range for \mbox{$\thicksim 500$} and \mbox{1000} actuators (DM correction range respectively to 11 and 16 \mbox{$\lambda/D$}): it shows the wavefront shaping gain as a function of initial speckle annulus frequency in \mbox{$\lambda/D$} for \mbox{1000} (black curve) and \mbox{500} (red curve) actuators. We see, in Fig.~\ref{fig:act_number0}, a large performance improvement when increasing the number of actuators illustrating the impact of DM actuator number inside the DM correction range. \\
Restraining the dark hole size and thus the needed correction range is a way to be less sensitive to this limitation as illustrated in \mbox{Fig.~\ref{fig:act_number1}} which shows the gain when doubling the actuator number for dark hole size of \mbox{10 $\lambda/D$} (black curve) and \mbox{5 $\lambda/D$} (red curve). The gain when doubling the actuator number is defined as
\begin{equation}
\mathrm{
\Gamma = \frac{\gamma_{1000}}{\gamma_{5 00}}}, 
\end{equation}
where $\gamma_{N}$ is the wavefront shaping gain (defined in equation \ref{eq:gain}) with \textit{N} actuators. We see that doubling the actuator number has less of an impact on the gain in performance for the smallest dark hole. 


\subsection{Aliased speckles}
\label{sec:folding}
\begin{figure}
\centering
\includegraphics[height=2.3 cm]{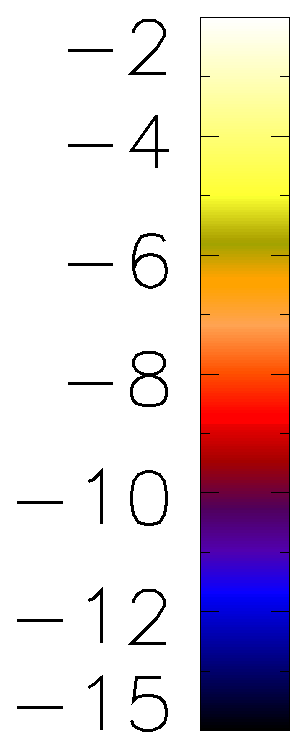}
\includegraphics[height=2.3 cm]{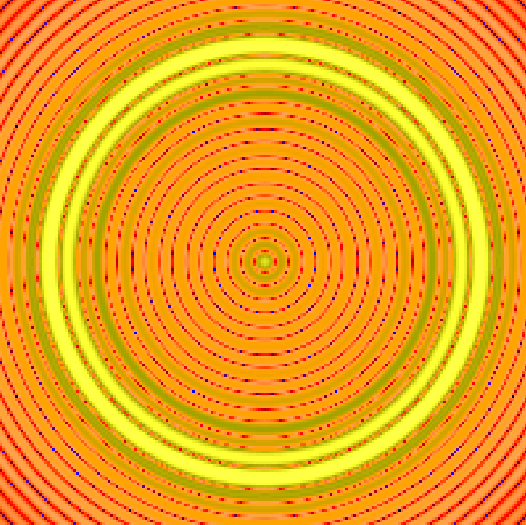}
\includegraphics[height=2.3 cm]{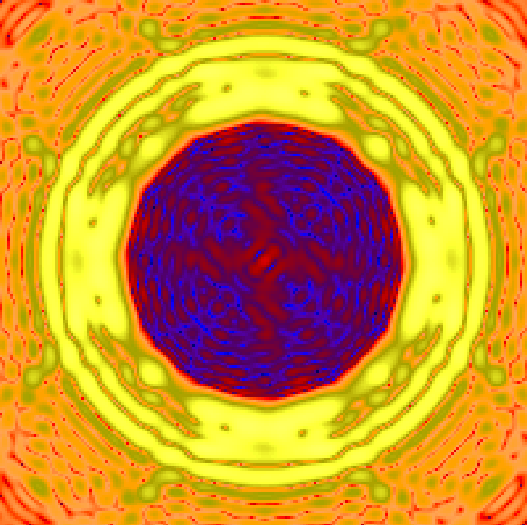}
\includegraphics[height=2.3 cm]{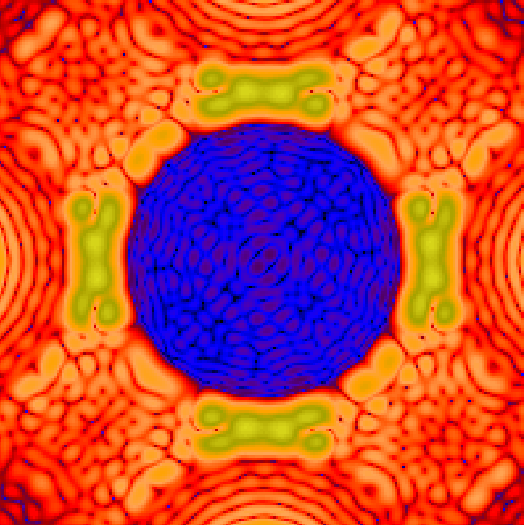}
\caption{Initial annulus of speckles at frequency outside the deformable mirror correction range (left), after the dark hole algorithm (middle), and the dark hole algorithm result when applied with an annulus of speckle at a frequency inside the deformable mirror correction range. The scale is logarithmic.}
\label{fig:aliasedill}
\end{figure}

\begin{figure}
\centering
\includegraphics[width=1.\columnwidth]{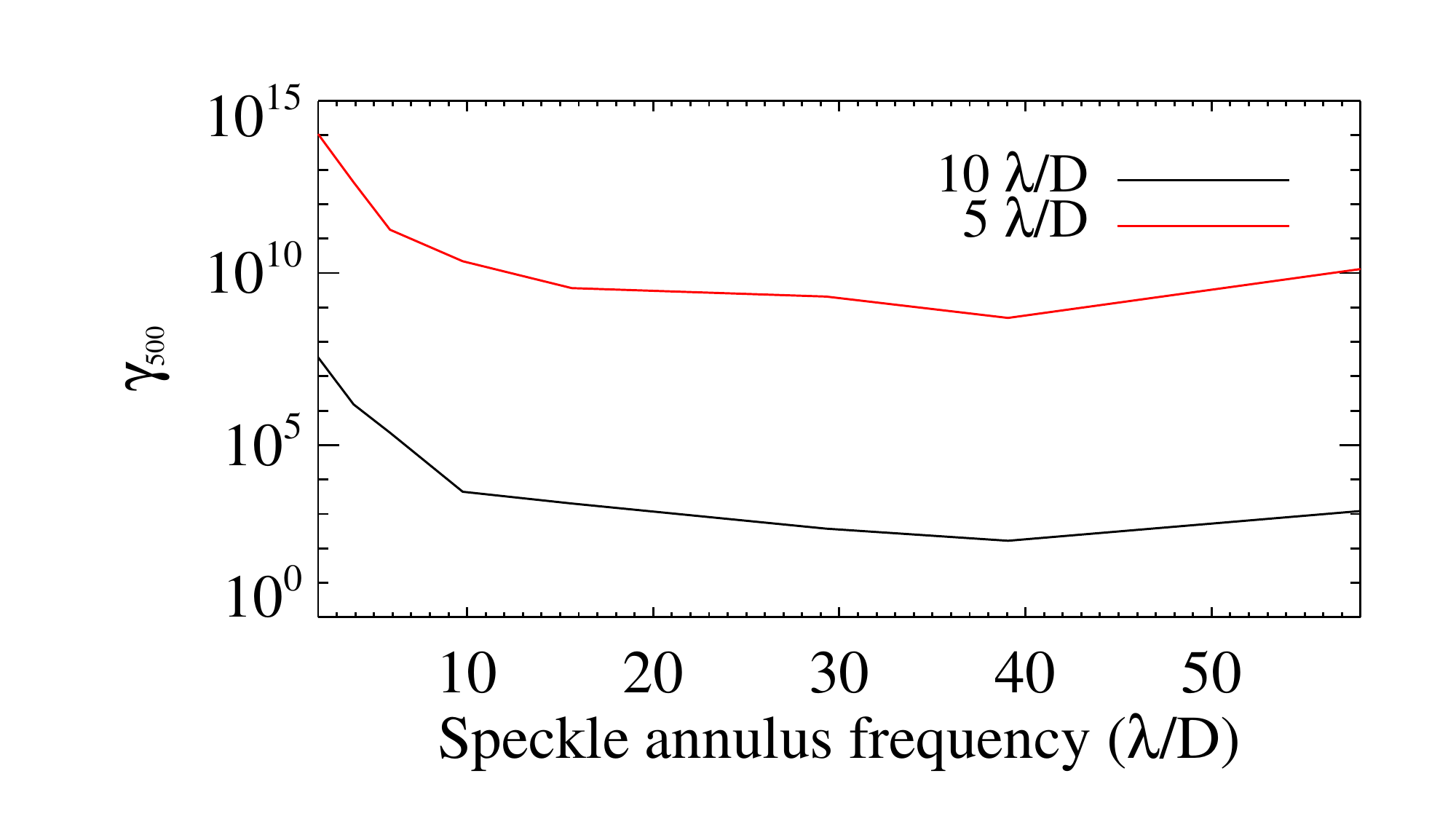}
\caption{Gain in contrast with 500 actuators ($\gamma_{500}$) when computing the wavefront shaping algorithm for speckles frequencies inside and outside a dark hole of size \mbox{10 $\lambda$/D} (black) and \mbox{5 $\lambda$/D} (red).}
\label{fig:folding0}
\end{figure}

Aliased speckles are speckles that are present in the dark hole owing to the diffraction pattern convolution. Those aliased speckles can be evidenced by assuming a small phase and approximating the complex amplitude at the pupil plane with a Taylor series, up to the first order, as
\begin{equation}
\mathrm{
E_{p}=A e^{i\varphi} \approx A(1+i\varphi).}
\end{equation}
We then introduce a perfect coronagraph, which removes the constant term in the pupil plane (derived from \citealt{CavarrocBoccalettiBaudozEtAl2006}) such that $\mathrm{E_{p}=iA\varphi}$. If we assume a Fourier transform relationship between the pupil and image planes, the electric field at the focal plane can be written as
\begin{equation}
\mathrm{
E_f=i} \widehat{\mathrm{A\varphi},}\\
\label{eq:linear}
\end{equation}
and the corresponding intensity as
\begin{equation}
I_{\mathrm{f}}=E_\mathrm{f} E_\mathrm{f}^* = |\widehat{A\varphi}|^2.
\label{eq:intensity}
\end{equation} 
The image plane intensity is the power spectrum of $A \varphi$ and degrades the high-contrast performance as it creates structures in the dark hole in the form of Airy rings. For instance, a pupil plane cosine phase pattern $\varphi$ at a frequency defined outside the dark hole creates at the focal plane, a wide lobe outside the dark hole but secondary halo lobes within the dark hole as the result of the convolution of the high-frequency peak with the Airy diffraction pattern (more generally, a combination of random frequencies will appear as a speckled halo). The characteristic size of these \textit{aliased speckles} is less than \mbox{$1 \lambda/D$}, as they originate from the combination of multiple halo ring structures. Therefore they can only be partially corrected with energy minimization. This impact is reduced when using an apodized coronagraph as, in this case, $\hat{A}$ decreases rapidly with frequency and limits the aliased speckle intensities \citep{GiveOnKasdinVanderbeiEtAl2006}. \\
In order to estimate the degradation resulting from these aliased speckles, we apply the dark hole algorithm to different initial phase patterns (radial cosine at different frequencies as described in Section \ref{sec:act_number}), when assuming the focal plane electric field as defined in equation \ref{eq:linear}. \mbox{Fig.~\ref{fig:aliasedill}} shows, at the same logarithmic scale, the initial annulus of speckles at a frequency outside the DM correction range (left), its corresponding pattern after the dark hole algorithm (middle), and the dark hole algorithm result with an annulus of speckles at a frequency inside the DM correction range (right). We illustrate here the DM limitation to correct for speckles outside the DM correction range (residual energy inside the dark hole in the middle of the figure) that is not present when the initial aberration frequency is inside the correction range (right in figure). 
\\
The black curve in Fig.~\ref{fig:folding0} shows the algorithm capability to correct for frequencies inside and outside the dark hole. It is the gain owing to wavefront shaping (from equation \eqref{eq:gain}) with a DM with \mbox{$\thicksim 500$} actuators (DM correction range up to \mbox{11 $\lambda / D$}) within a dark hole of width \mbox{10 $\lambda/D$}. The algorithm performance decreases as the phase pattern frequency increases (1) inside the dark hole owing to the DM inability to reproduce higher-frequency pattern well (see Section \ref{sec:act_number}) and (2) outside the dark hole owing to aliased speckles. \\
One way to minimize this effect is to decrease the dark hole size \citep{BordeTraub2006}, such that the algorithm is able to mimic similar pattern speckles with a central frequency outside the dark hole area but inside the DM correction range.
The red curve in \mbox{Fig.~\ref{fig:folding0}} shows the gain owing to wavefront shaping when the dark hole size is decreased to \mbox{5 $\lambda/D$}. Decreasing the dark hole size improves the high-contrast performance (1) for frequencies outside the dark hole, as the algorithm mimics aliased speckles by putting more energy at frequencies outside the dark hole area but also (2) for frequencies inside the dark hole: as the dark hole size decreases, the needed DM frequencies decrease such that the DM better reproduces the phase pattern at those frequencies (which is equivalent to increasing the number of actuators as in Section \ref{sec:act_number}).

\subsection{Summary}
In this section, we analytically or semi-analytically studied some limitations to high-contrast imaging. The actuator number and frequency folding significantly impact the performance, but this impact can be minimized by narrowing the dark hole. Furthermore, analytical analysis of the impact of the DM location with simplifying assumptions showed that high-contrast imaging at small separation requires a large setup and a small dark hole size. We now use a more realistic model to validate the impact of these limitations.

\section{Model definition}
\label{sec:model}
This section defines the model we use to simulate high-contrast imaging at a small separation with two DMs. In the following, we do not treat (1) quasi-static aberrations, as we assume a correction with a timescale shorter than structural or thermal changes, or (2) dynamical aberrations. The second assumption can be realized with a spatial instrument or by assuming that atmospheric turbulence has been corrected by an ExAO system (see further comments in the \textit{Discussion} section.
\subsection{Numerical assumption}
We assume a perfect coronagraph that removes all the coherent light without aberration. The light is propagated along a setup free of aberration, up to a \textit{coronagraphic pupil plane}, where the complex amplitude is recorded and subtracted when running simulations with aberrations. We thus assume that our perfect coronagraph is not able to correct any term owing to aberration.
For small aberration, the pupil electric field is defined as
\begin{equation}
\mathrm{
E_{p}=A+  iA \varphi -  \frac{1}{2} A \varphi^2.} 
\end{equation}
The perfect coronagraph removes the deterministic (constant) term but cannot correct for the linear term \mbox{$\mathrm{iA \varphi}$} nor for the quadratic amplitude term \mbox{$\mathrm{\varphi^2 / 2}$} (the perfect coronagraph defined here is sensitive to the aberrations and thus cannot correct for the phase contribution). Some perfect coronagraph definitions in the literature \citep{CavarrocBoccalettiBaudozEtAl2006,SauvageMugnierRoussetEtAl2010} minimize the integrated energy after a coronagraph (by subtracting the mean of this quadratic term \mbox{$\mathrm{\varphi^2 / 2}$}) and thus better reflect a coronagraph that removes the central pattern (e.g a four-quadrant phase mask or vortex). Because our simulation attempts to define the ability to correct for amplitude and phase aberrations with two DMs when using different phase patterns, we choose to define a coronagraph sensitive to aberrations. \\
Furthermore, we focus in this paper on the linear approach of the algorithm that minimizes the overall energy inside the dark hole (Section \ref{sec:energy}). The DM coefficients are computed using equation \eqref{eq:coeff}. We do not restrain the DM stroke (EFC or stroke minimization) because we want to assess the impact of setup parameters, especially the DM distances, on high-contrast performance regardless of the stroke values. In practice, the DM strokes in our simulation are within the algorithm linear regime. We finally assume a perfect estimation of the complex amplitude at the focal plane.

\subsection{Numerical implementation}
\subsubsection{Optical model} 
\label{opt_mod}
An end-to-end model is defined based on a generic setup for high-contrast imaging. The setup consists of \mbox{$\thicksim$ 25} optics (which is typical of current high-contrast imaging instruments) containing, amongst other items, a pupil simulator, two DMs and a perfect coronagraph. Paraxial lenses ensure the transition between the image and pupil planes. The setup is monochromatic at \mbox{1.65 $\mu$m}, and the pupil is assumed to be circular with a diameter of \mbox{7.7 mm}. \\
Active optics (DMs) have 32 $\times$ 32 actuators with 300-\mbox{$\mu$m} pitch, corresponding to 22 $\times$ 22 active actuators in the pupil. The optical baseline allows the testing of DM distances up to \mbox{2.5 m}. \\
Passive optics have a circular shape amplitude, assumed to be larger than the pupil diameter (four times the pupil diameter). Each optic is computed with random static aberrations defined by their total amount of aberration (in nm rms) and their frequency distribution (power law of the power spectral density, PSD). For statistical analysis, \mbox{128} phase realizations are defined per optic.
We do not add amplitude error on optics, such that the amplitude error present in the setup results from the Fresnel propagation of phase errors.
 
\subsubsection{Dark hole algorithm implementation} 
\begin{figure}
\centering
\includegraphics[width=.98\columnwidth]{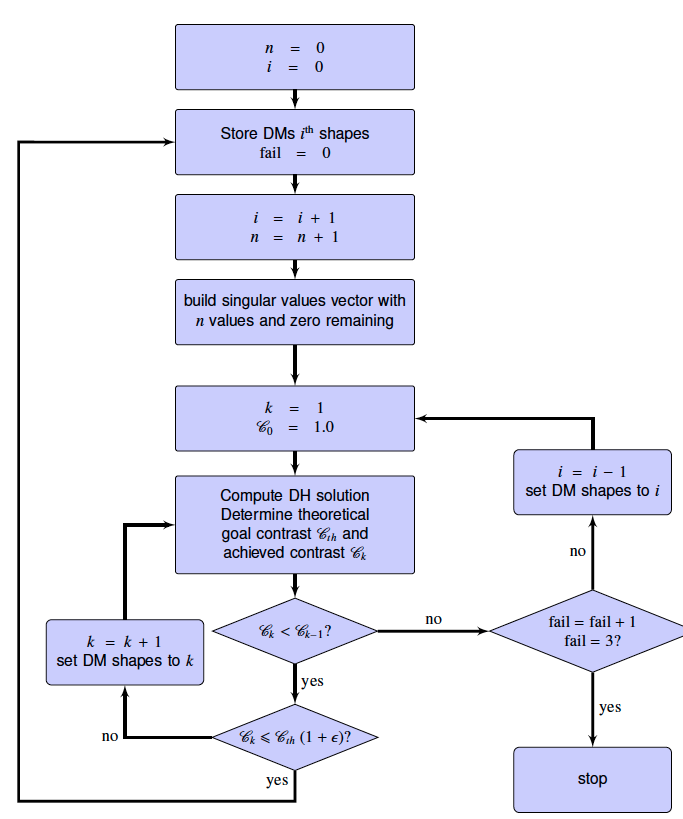}
\caption{Flow diagram of the dark hole algorithm.}
\label{fig:flowdiagram}
\end{figure}

The interaction matrix $\mathrm{M_0}$ from equation \eqref{eq:coeff} defines the system response to each DM actuator and is numerically computed by poking each DM actuator (with phase \mbox{$\ll$ 1 rad} to maintain the linear approximation), then propagating along the setup and recording the complex amplitude at the image plane. We note that the computation of $\mathrm{M_0}$ assumes no aberration on the optics, which is equivalent, for a setup with low noise level, to poking the actuator by positive/negative values to remove the aberration contribution. The real part of the matrix is inverted using singular value decomposition (SVD), and the singular value vector is sorted from the highest to the lowest values. In order to avoid divergence and optimize the algorithm, we use a two-level iterative process, as illustrated in figure \ref{fig:flowdiagram}. The first iterative process impacts the singular value vector, which is built up by taking the \textit{n} first singular values and zeroing the remaining. The dark hole solution (DM coefficients) is then computed in a second iterative process, until the obtained contrast is close to the theoretical one (by a factor of \mbox{$1+\epsilon$}, with $\epsilon$ typically of the order of 10\%). The algorithm then adds the next $n$ singular values (that was zeroed in the previous step) to the singular vector, and this process is repeated until an overall best contrast is obtained. If a better contrast cannot be reached, the previous DM shapes are used as a starting point to the process : this prevents the algorithm from stopping when close singular values give similar performance and thus prevent a local minimum solution.
\\ The singular value threshold \textit{n} at each iteration is empirically determined as a compromise between the performance and the computational time and depends mainly on the setup aberration level. \\
High-contrast imaging around \mbox{$1 \lambda/D$} requires a large setup (DM distances of a few meters) and a small dark hole as described in Section \ref{dmloc}. We thus define for the simulation a dark hole from 0.8 to \mbox{4 $\lambda/D$} to emphasize performance at very small separations.

\subsubsection{Numerical code} 
The code we use for the Fresnel propagation between each optical element is \textsc{proper} \citep{Krist2007a}. This code uses the angular spectrum and Fresnel approximation as propagation algorithms; the procedure automatically determines which is the best algorithm to implement. \textsc{proper} and the dark hole algorithm were written in \textit{IDL} but ported to \mbox{C++}, such that the computation of several configurations can be performed simultaneously in a data center available at Observatoire de la C\^{o}te d'Azur to speed up the computational time (from one day to several hours).
The numerical pupil diameter size is \mbox{400} pixels for a grid size of \mbox{2048} pixels.

\section{Numerical results}
 \label{sec:results}
 \subsection{Illustration of simulated dark holes}
\begin{figure*}
\centering
\includegraphics[height=4. cm]{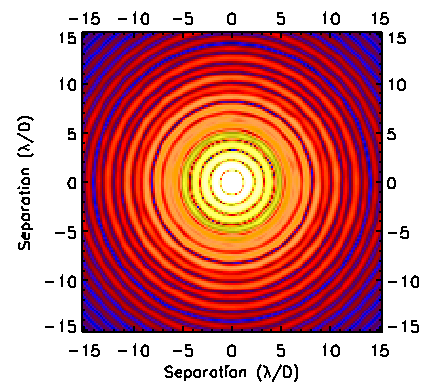}
\includegraphics[height=4. cm]{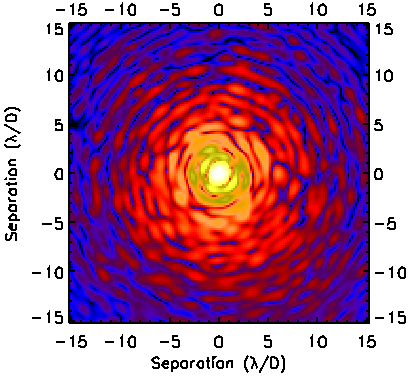}
\includegraphics[height=4. cm]{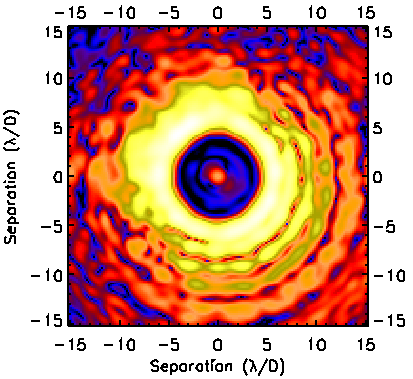}
\includegraphics[height=4. cm]{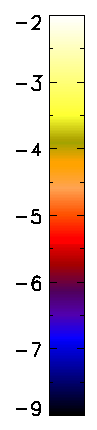}
\caption{Contrast ratio image (logarithmic scale) before (left) and after (middle) the coronagraph and after the wavefront shaping with the two deformable mirrors (right)}
\label{fig:image}
\end{figure*}

For illustration purposes, this section describes the high-contrast capability of our numerical dark hole implementation and the metric used to estimate performance. We focus on the case where $\mathrm{DM_1}$ is in the pupil plane and $\mathrm{DM_2}$ is \mbox{0.5 m} from the pupil plane. This case is not the DM location that provides the best performance but solely represents an example of achievable contrast. Fig.~\ref{fig:image} shows the image contrast ratio (logarithmic scale) before the coronagraph (left), after the coronagraph (middle) and after the wavefront shaping (right). The contrast ratio used in the following is the one defined in Section \ref{sec:act_number}. The setup optic aberrations are set to 5 nm rms per optic (overall setup aberration level of about 20 nm rms), with the aberration PSD in $f^{-3}$, where \textit{f} is the spatial frequency. 
The metric used in the following to estimate the high-contrast imaging performance is the \mbox{5$\sigma$} contrast ratio $\mathscr{C}$.

\subsection{Impact of DM location on high-contrast performance}
\label{simu_dmdist}
\begin{figure}
\centering
\includegraphics[width=.95\columnwidth]{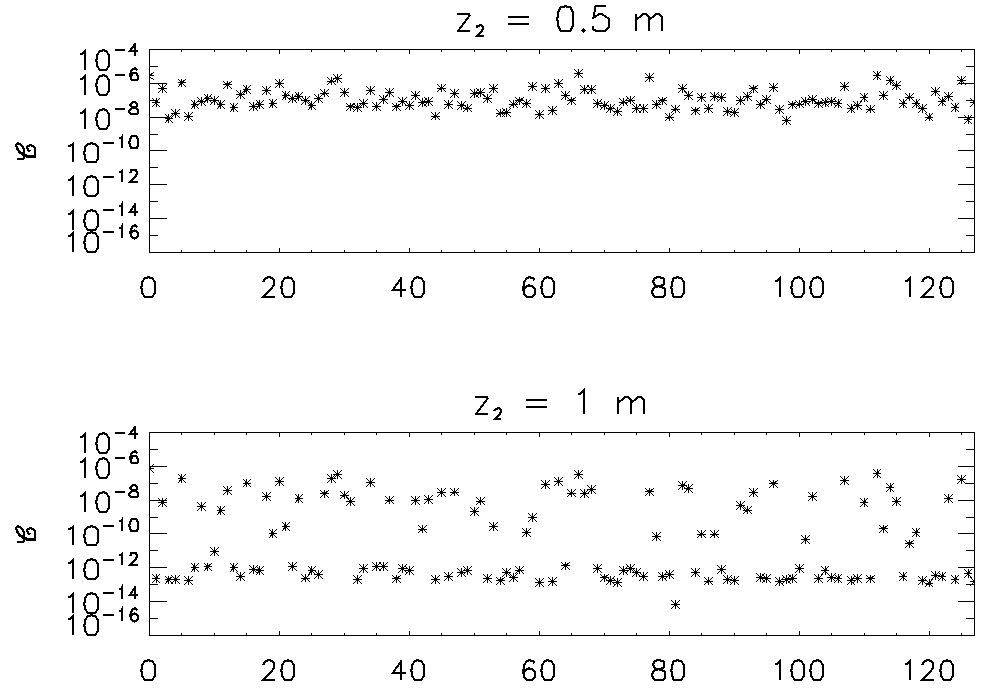}
\includegraphics[width=.95\columnwidth]{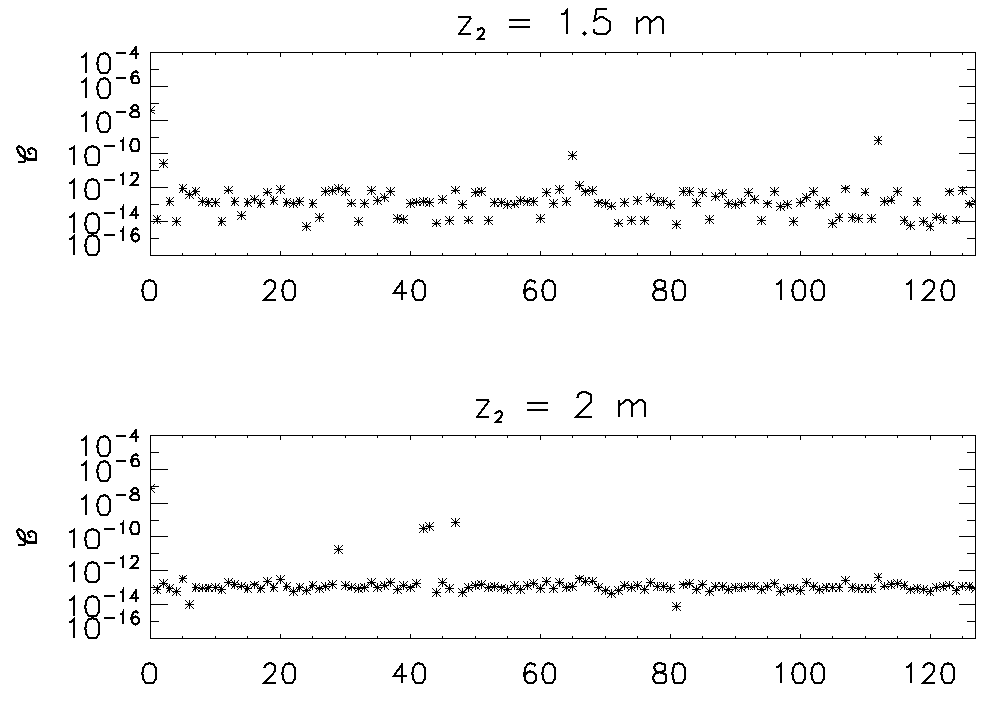}
\includegraphics[width=.95\columnwidth]{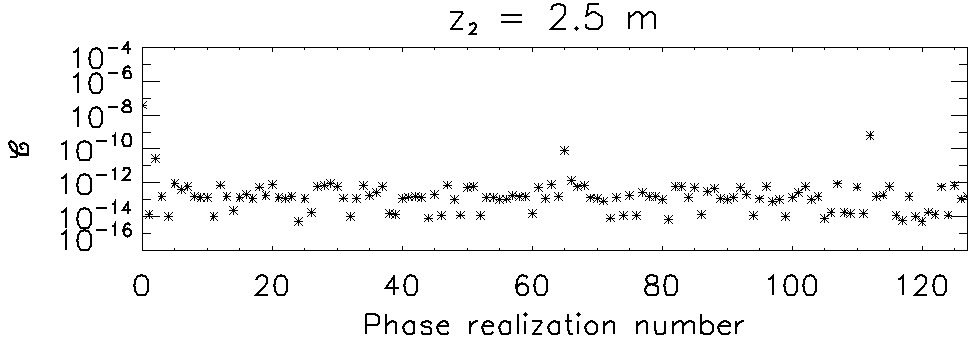}
\caption{5$\sigma$ contrast ratio within a dark hole (from 0.8 to \mbox{4 $\lambda$/D}) as a function of the phase realization number for $\mathrm{DM_1}$ in the pupil plane and $\mathrm{DM_2}$ situated (from top to bottom) 0.5, 1, 1.5, 2 and 2.5 m from the pupil plane.}
\label{fig:dm0}
\end{figure}
\begin{figure*}
\centering
\includegraphics[width=2.1\columnwidth]{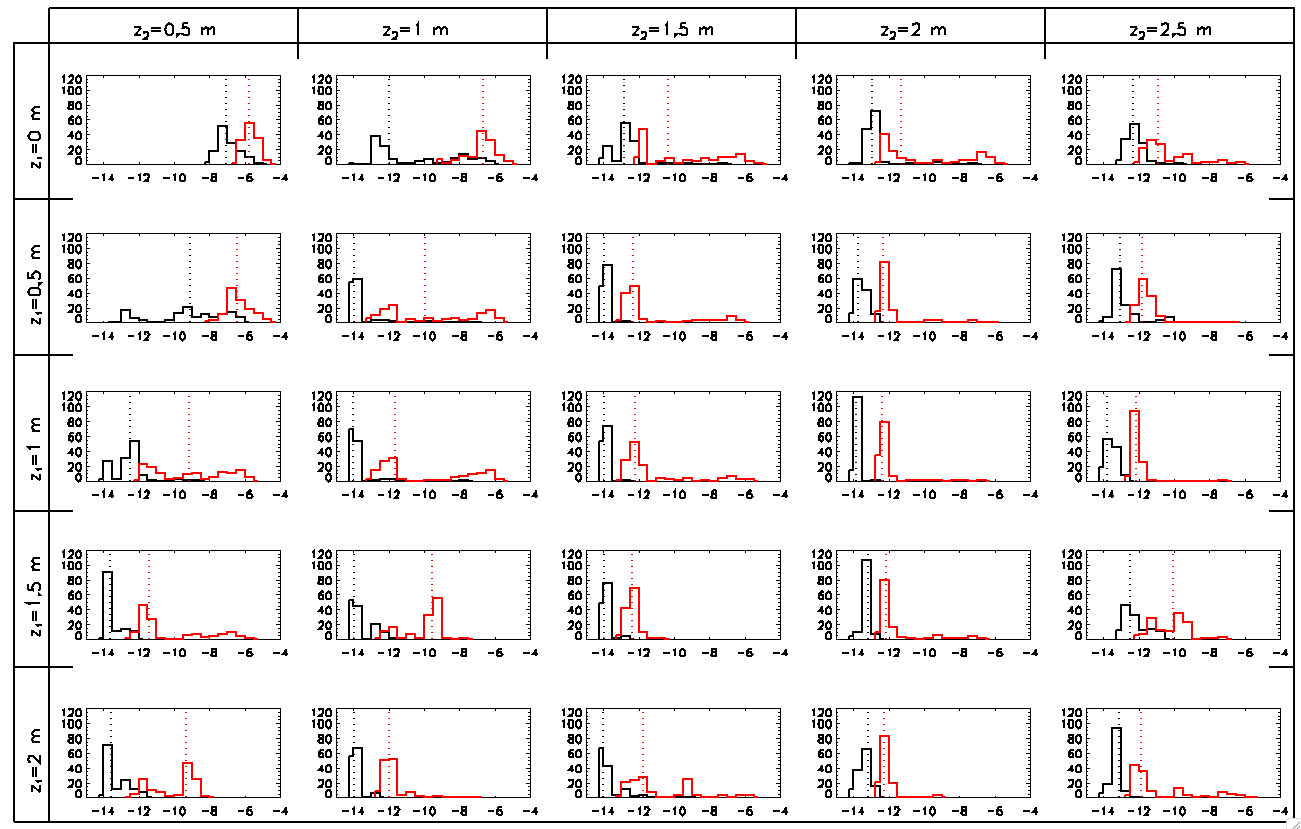}
\caption{5$\sigma$ contrast ratio histogram for each DM location. The histogram is the number of realizations that achieve the contrast ratio defined on the abscissa (scale in powers of ten). The black and red histograms correspond to setup aberrations of 20 and 40 nm rms. The dotted lines are the contrast median for each case.}
\label{fig:dm1}
\end{figure*}

We assume that each optic contains an aberration of 5 nm rms and a PSD power law in \mbox{$f^{-3}$} which is typical of current manufacturing processes. For statistical analysis, we computed 128 different random aberration realizations for each optic in the setup.\\
\mbox{Figure \ref{fig:dm0}} is an illustration of the typical output that will be used in the forthcoming sections. Each asterisk represents the \mbox{5$\sigma$} contrast ratio median computed within the dark hole (defined from 0.8 to \mbox{4 $\lambda$/D}), when $\mathrm{DM_1}$ is in the pupil plane and $\mathrm{DM_2}$ is at various distances from the pupil plane. As $z_2$ increases, the overall contrast ratio decreases, reaching a minimum of about $10^{-13}$ when $\mathrm{DM_2}$ is \mbox{$\thicksim$ 2 m} from the pupil plane and increasing for the largest distance (\mbox{2.5 m}). This result is consistent with the semi-analytical analysis of Section \ref{dmloc} with an optimum $z_2$ around \mbox{1.8 m} when the first DM is in pupil plane. We note a large dispersion in phase realization when the DM distance is not optimum ($z_2$ of \mbox{1 m} for instance). The algorithm achieves a very low  contrast ratio ($10^{-13}$) because the coronagraph is assumed to remove all the light without aberration and because we take into account only amplitude errors from the propagation of static phase errors, so that the fundamental limitation at small separation can be assessed.  \\
The DM distances are tested in several cases: with $\mathrm{DM_1}$ in the pupil plane, at 0.5, 1, 1.5 and \mbox{2 m} from the pupil plane, and with $DM_2$ at 0.5, 1, 1.5, 2 and \mbox{2.5 m} from the pupil plane. The contrast ratio histograms for each DM location (when varying $z_1$ and $z_2$) are presented in \mbox{Fig.~\ref{fig:dm1}}. Each plot represents the number of realizations (ordinate) that reaches a given \mbox{5$\sigma$} contrast (abscissa, in logarithm scale). Black and red histograms represent simulations with respectively 5 and 10 nm rms per optic and with PSD in $f^{-3}$ (overall setup amount of \mbox{$\thicksim$20} and \mbox{$\thicksim$40 nm rms})\footnote{The 128 realizations for each distance use the same 128 sets of aberrations (same 128 \textit{seeds})}. The dotted line is the median of the achieved contrast ratio for each case. For illustration, the plot in \mbox{Fig.~\ref{fig:dm0}} is represented by the first row in \mbox{Fig.~\ref{fig:dm1}}. As each DM location increases, histograms are sharper (less dispersion) and the medians (dotted lines) are lower (better contrast). Optimum performance with 20 nm rms is obtained for $z_2$ between1.5 m and 2 m when \mbox{$z_1$=\{0, 0.5, 1 m\}}, for \mbox{$z_2$ = 1.5 m} when \mbox{$z_1$ = 1.5 m} and for \mbox{$z_2$ = 1 m} when \mbox{$z_1$ = 2 m}, with a slightly higher contrast at \mbox{$z_1$ = $z_2$ = 1.5 m}. These results are in agreement with the analytical approach (see Fig.~\ref{fig:dmlocation1}), illustrating the fact that the approximation used in Section \ref{dmloc} is valid to estimate the optimum DM distances. Increasing the amount of aberration degrades the overall performance (because the setup is outside of the linear assumption used in the energy minimization algorithm) but does not impact the optimum DM location. Plots on Fig.~\ref{fig:dmlocation1} also shows large bimodal histograms in several cases (for instance when \mbox{$z_1$ = 0.5 m} and \mbox{$z_2$ = 1.5 m}), illustrating when the algorithm is out of the linear assumption but also the fact that our dark hole numerical computation could be optimized with a more complex algorithm.

\subsection{Impact of aliased speckle (aberration PSD)}
\label{sec:alias}
Section \ref{sec:folding} showed for a simple case that speckles with frequencies outside the dark hole create aliased speckles inside the dark hole. In order to estimate the impact of these aliased speckles on the performance, we simulate an aberration pattern with a greater PSD power-law exponent than in the previous section. The PSD power law impacts the aberration frequency distribution: the amount of aberration at large frequencies increases when the power law increases. We test two power-law realizations: with less aberration at high frequencies, and thus less aliased speckles (PSD in $f^{-3}$), and with more aberration at high frequencies (PSD in $f^{-2.5}$). 
\begin{figure}
\centering
\includegraphics[width=\columnwidth]{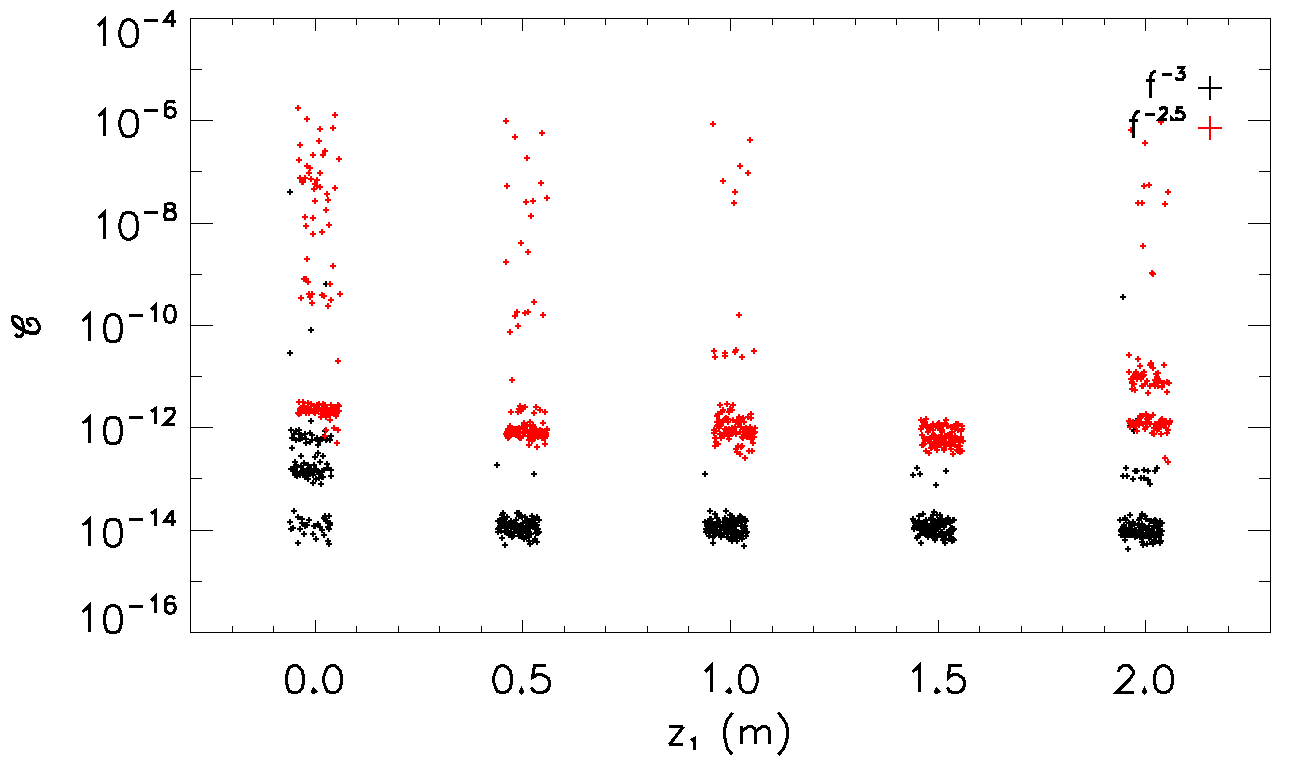}
\caption{5$\sigma$ contrast ratio within the dark hole (from 0.8 to \mbox{4 $\lambda$/D}) as a function of $z_1$ for \mbox{$z_2$ = 1.5 m}. Each plus sign corresponds to a random phase realization. Black and red plus signs represent the realization with the power spectral density in respectively $f^{-3}$ and $f^{-2.5}$.}
\label{fig:psd0}
\end{figure}
The simulation is realized in a representative case: with $\mathrm{DM_2}$ at \mbox{1.5 m} from the pupil plane and with $\mathrm{DM_1}$ at the pupil plane and at 0.5, 1, 1.5 and \mbox{2 m} from the pupil plane. This case is representative because the contrast increases as $z_1$ increases, with good performance for $z_1$ between 0.5 and \mbox{1.5 m} (see column 4 in Fig.~\ref{fig:dm1}). The results are presented in Fig.~\ref{fig:psd0}, where the \mbox{5$\sigma$} contrast ratio within the dark hole is shown as a function of $z_1$ for a $z_2$ of \mbox{1.5 m}. Each plus sign is a phase realization, and black and red plus signs represent the computation with the PSD in $f^{-3}$ and $f^{-2.5}$ respectively. We note a significant performance degradation when increasing the PSD power law (red curve), with larger dispersion in phase realization and a worse contrast ratio. This is consistent with Section \ref{sec:folding}, which shows the large impact of aliased speckles in high-contrast imaging. The optimum $\mathrm{DM_1}$ location is unchanged for the two simulations (\mbox{1.5 m}).
\subsection{Impact of actuator number}
\begin{figure}
\centering
\includegraphics[width=\columnwidth]{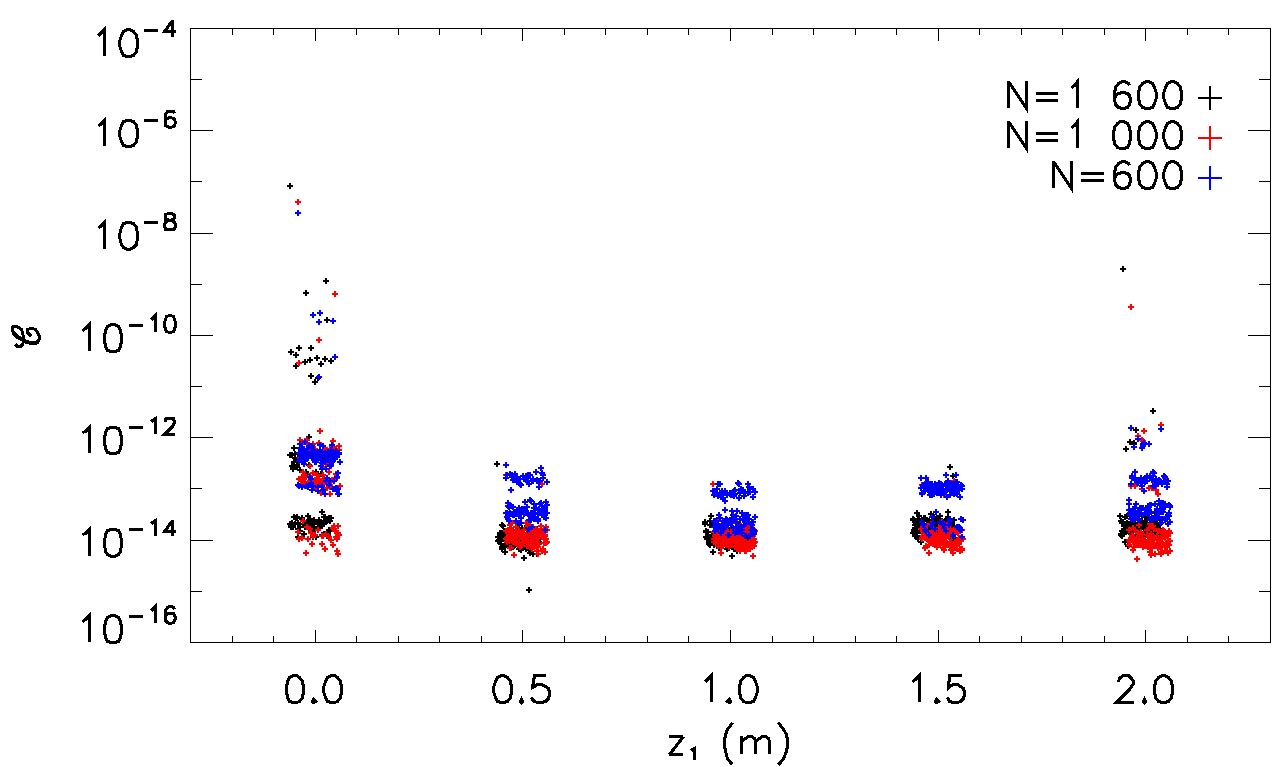}
\caption{5$\sigma$ contrast ratio within the dark hole (from 0.8 to \mbox{4 $\lambda$/D}) as a function of $z_1$ for \mbox{$z_2$ = 1.5 m}. Each plus sign corresponds to a random phase realization. Black, red and blue plus signs correspond to the simulation with respectively 1600, 1000 and 600 actuators.}
\label{fig:actnum}
\end{figure}

This subsection illustrates the impact of actuator number on high-contrast imaging in a representative case (same as Section \ref{sec:alias}), with a $z_2$ of \mbox{1.5 m} and \mbox{$z_1$=\{0, 0.5, 1, 1.5 and 2 m\}}. The number of actuators is tested with \mbox{24 $\times$ 24} (\mbox{$\thicksim$ 600}), \mbox{32 $\times$ 32} (\mbox{$\thicksim$ 1 000}) and \mbox{40 $\times$ 40} (\mbox{$\thicksim$ 1 600}) actuators per DM. The obtained \mbox{5$\sigma$} contrast ratio for $z_1$ from 0 to 2 m from the pupil plane is shown in Fig.~\ref{fig:actnum} (from top to bottom). The contrast with 600 actuators (black plus signs) is slightly worse than the contrast with \mbox{1000} and \mbox{1600} actuators, showing that the dark hole is sufficiently small (from 0.8 to \mbox{4 $\lambda$/D}) to be negiglibly impacted by the actuator number (see Section \ref{sec:act_number}).

\subsection{Impact of optical design architecture}
\begin{figure}
\centering
\includegraphics[width=.9\columnwidth]{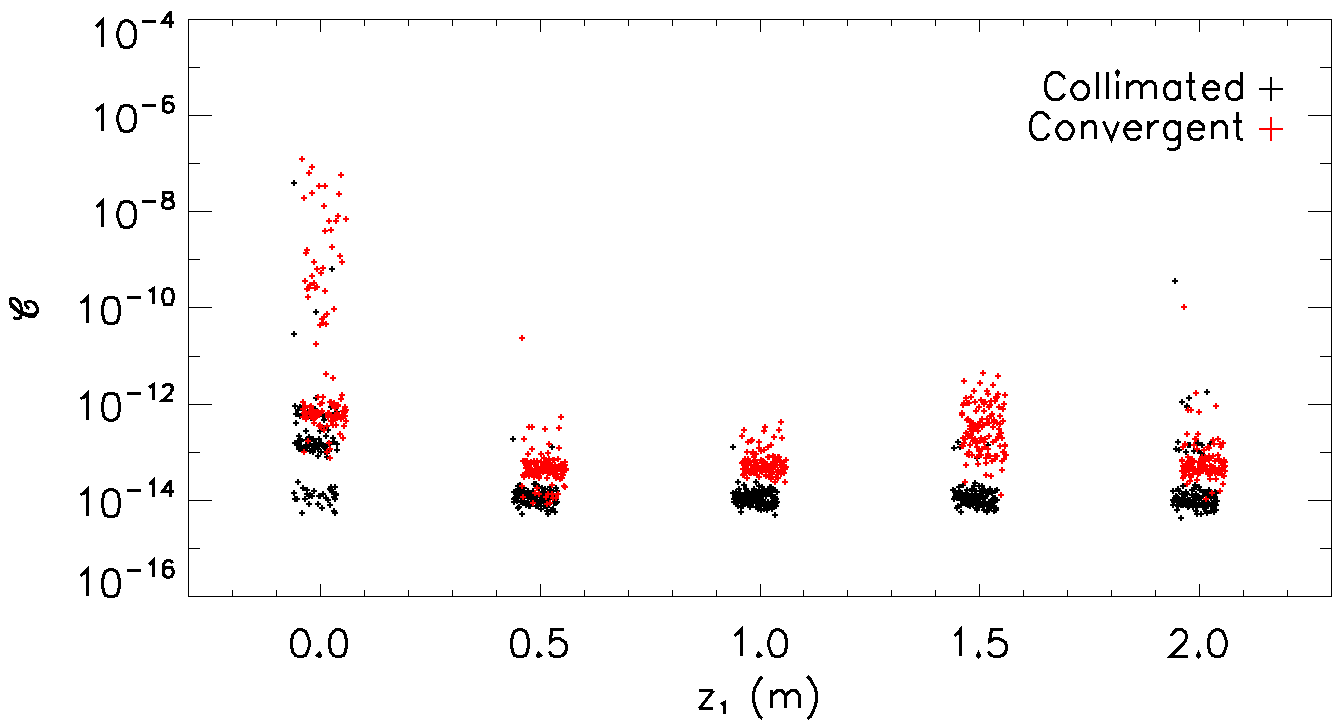}
\caption{5$\sigma$ contrast ratio within the dark hole (from 0.8 to \mbox{4 $\lambda$/D}) as a function of $z_1$ for \mbox{$z_2$ = 1.5 m}. Each plus sign corresponds to a random phase realization. Black and red plus signs correspond to the simulation in collimated and convergent beams, respectively.}
\label{fig:archi1}
\end{figure}

The impact of the optical design architecture is determined with $\mathrm{DM_2}$ at \mbox{1.5 m} from the pupil plane and $\mathrm{DM_1}$ at the pupil plane and at 0.5, 1, 1.5 and \mbox{2 m} from the pupil plane (same as in the previous subsections). We test two architectures: with $\mathrm{DM_2}$ in a collimated or a convergent beam (see Section \ref{archi}). The size of the beam at $\mathrm{DM_2}$ is 0.9 times the size of the corresponding collimated beam, leading to a small loss in DM actuators (\mbox{$\thicksim$20\% loss in a given area}). The results are presented in Fig.~\ref{fig:archi1}: each plot is the \mbox{5$\sigma$} contrast ratio as a function of $z_1$ for the setup in collimated (black plus signs) and convergent (red plus signs) beams. We see no significant impact (slight performance degradation) on the high-contrast performance with the two architectures, because the dark hole size is small and thus less sensitive to the actuator number. The two architectures are thus almost equivalent in terms of performance, as long as the loss in actuator number does not impact the performance.

\subsection{Impact of dark hole size}
\label{dhsize}
\begin{figure}
\centering
\includegraphics[width=\columnwidth]{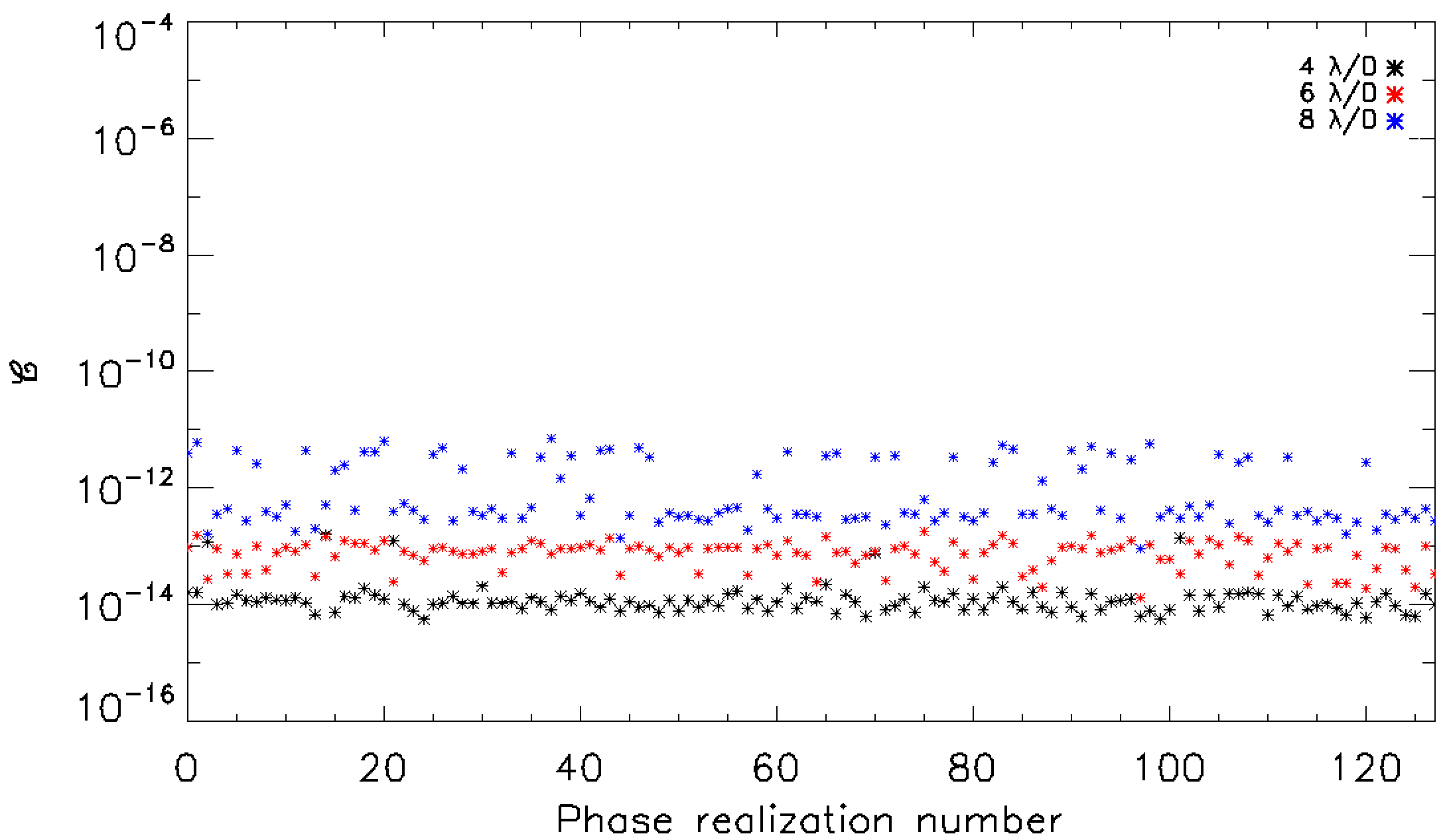}
\caption{5$\sigma$ contrast ratio as a function of the phase realization number for $\mathrm{DM_1}$ and $\mathrm{DM_2}$ at 1.5 m from the pupil plane. Black, red and blue asterisks correspond to the simulation for dark holes defined respectively from 0.8 to \mbox{4 $\lambda$/D}, from 0.8 to \mbox{6 $\lambda$/D} and from 0.8 to \mbox{8 $\lambda$/D}.}
\label{fig:dhsize}
\end{figure}
This subsection describes the impact of the dark hole size on the overall performance. The simulation is realized for the two DMs at \mbox{1.5 m} from the pupil plane. We define three dark hole sizes: from 0.8 to \mbox{4 $\lambda$/D}, from 0.8 to \mbox{6 $\lambda$/D} and from 0.8 to \mbox{8 $\lambda$/D}. The results are presented in Fig.~\ref{fig:dhsize}. The figure shows the \mbox{5$\sigma$} contrast ratio as a function of phase realization number for the different dark hole sizes. As the dark hole size increases, the performance becomes worse and the dispersion increases, consistent with the analysis in Section \ref{dmloc}.

\section{Discussion}
\label{sec:discussion}
We defined the potential limitations when searching for high-contrast images at very small separations (about \mbox{1 $\lambda$/D}). An analytical or semi-analytical analysis with simple assumptions (Section \ref{sec:limitation}) shows that aliased speckles and actuator number significantly limit high-contrast imaging but that the effects can be mitigated by using small dark hole sizes. The analysis also shows that wavefront shaping at small separations with two DMs requires large DM setup distances and a small dark hole size owing to the modulation of out-of-pupil DMs. In depth end-to-end simulation is developed to take into account the Fresnel propagation of phase errors. The results show a significant performance dependence on the DM location (Section \ref{simu_dmdist}), on the aberration amount and the PSD power law owing to aliased speckles (Section \ref{sec:alias}), and on the dark hole size (Section \ref{dhsize}). 
\begin{figure}
\centering
\includegraphics[height=.5\columnwidth]{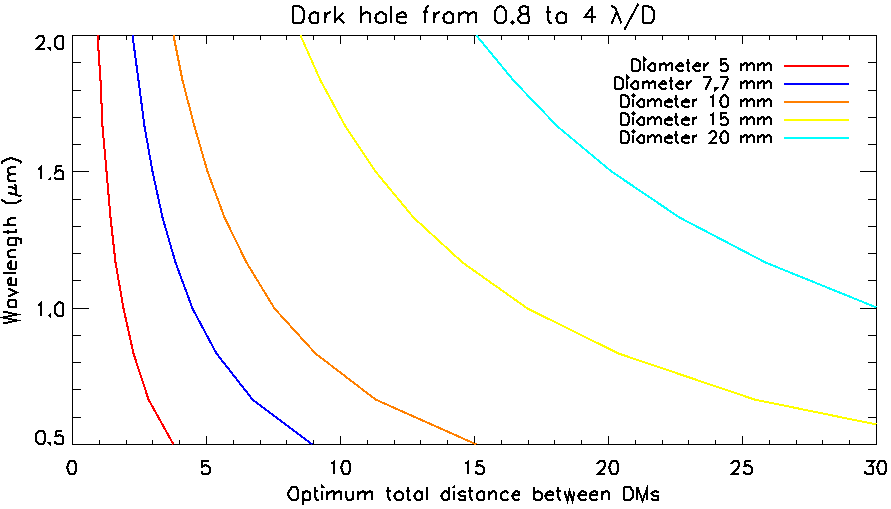}
\includegraphics[height=.5\columnwidth]{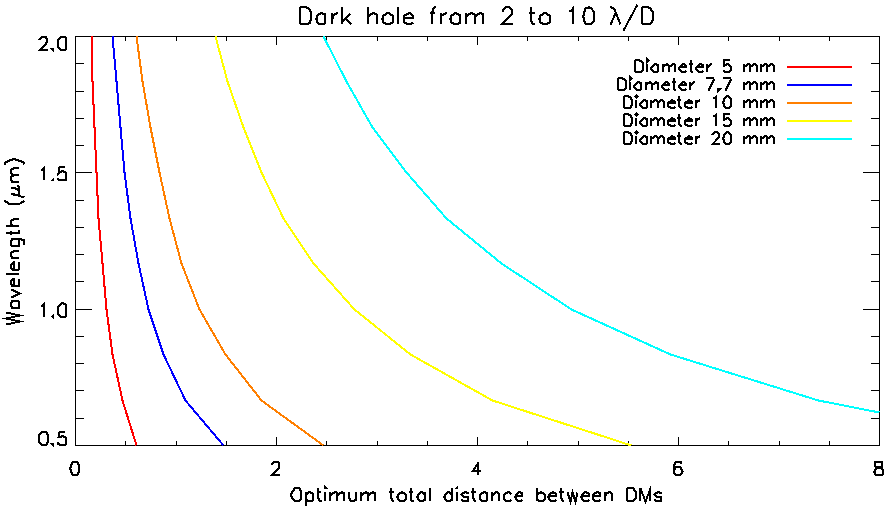}
\includegraphics[height=.5\columnwidth]{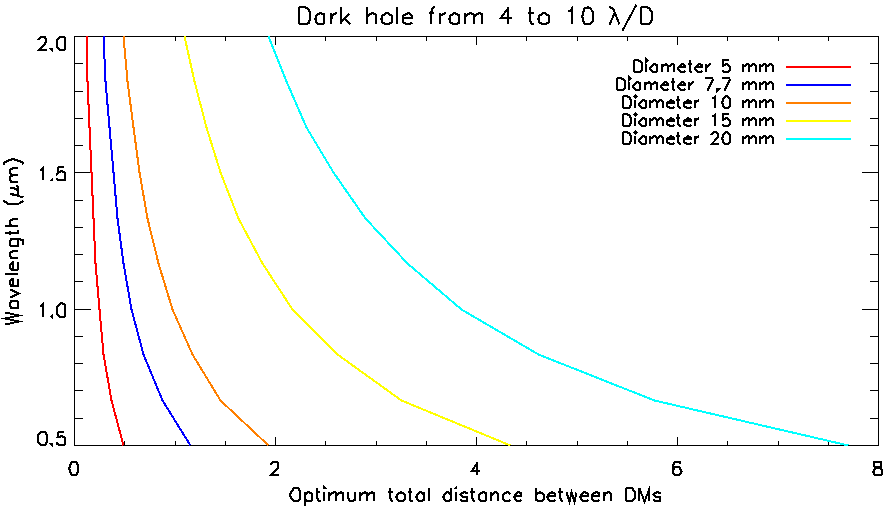}
\caption{Optimum distance between the two deformable mirrors for different dark hole sizes, wavelengths and pupil diameters. From top to bottom, the dark hole changes from 0.8 to 4, from 2 to 10 and from 4 to \mbox{10 $\lambda$/D}. Each plot represents the setup wavelength as a function of the optimum deformable mirror distance. The red, dark blue, orange, yellow and light blue curves represent respectively pupil diameters of 5, 7.7, 10, 15 and \mbox{20 mm}}
\label{fig:discuss}
\end{figure}
\begin{figure}
\centering
\includegraphics[width=.97\columnwidth]{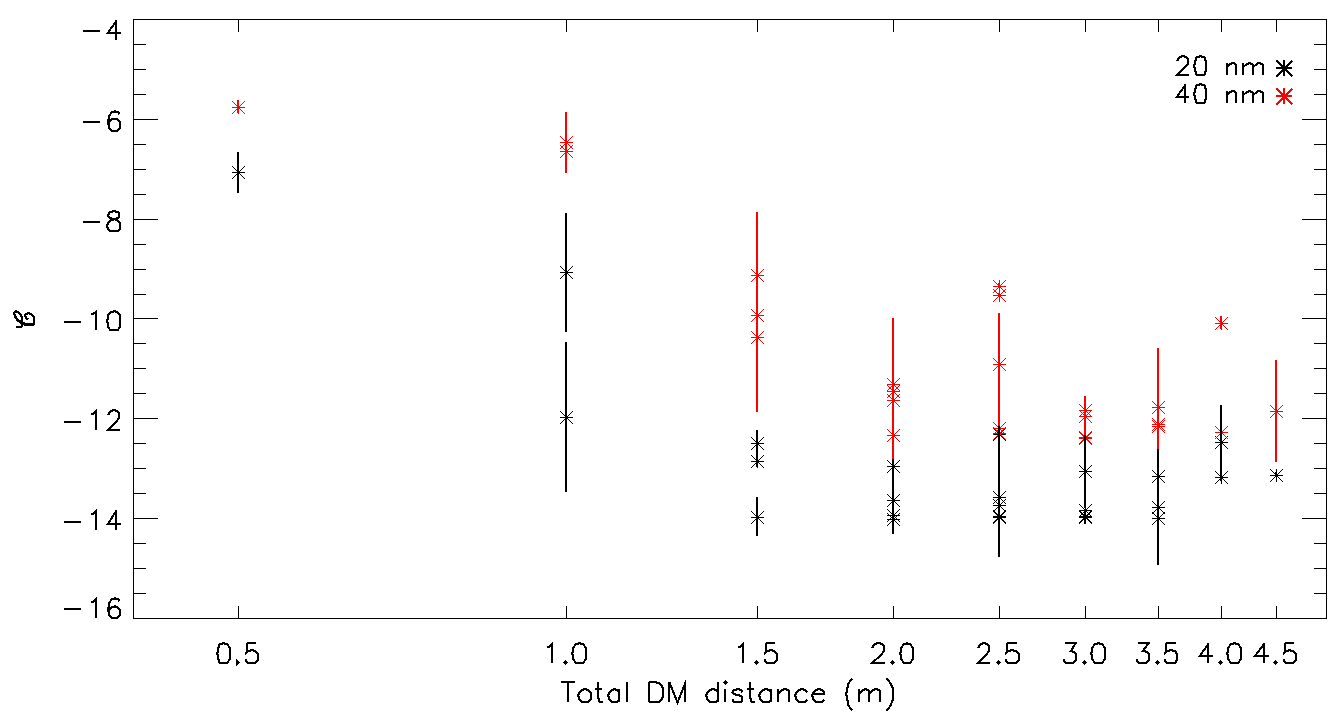}
\caption{Contrast ratio (asterisks, logarithmic scale) as a function of the total distance between the DMs for aberration amount of 20 nm (black) and 40 nm (red). Solid lines correspond to the dispersion (number of random realization that reaches contrast ratio greater that \mbox{5 $\sigma$} of the contrast values).}
\label{fig:discuss1}
\end{figure}
A PSD in $f^{-2.5}$ with \mbox{5 nm} per optic (overall setup amount of \mbox{$\thicksim$20 nm rms}) significantly degrades the contrast compared with a PSD in $f^{-3}$. Furthermore, the architecture and the actuator number do not significantly impact the results because the dark hole size is small. The optimum total distance between the two DMs can be estimated by the analytical approach, by defining the optimum distance as the one that equalizes the DM sine and cosine contributions at the focal plane (see Section \ref{dmloc}). Figure \ref{fig:discuss} shows the optimum total DMs distance for several setup parameters (pupil size and wavelength) and dark hole frequencies. It represents the optimum total distance between the two DMs ($z_1+z_2$) on the horizontal axis as a function of the wavelength (vertical axis) for (from top to bottom) the dark hole defined from 0.8 to \mbox{4 $\lambda$/D}, from 2 to \mbox{10 $\lambda$/D} and from 4 to \mbox{10 $\lambda$/D}. The red, dark blue, orange, yellow and light blue curves (plotted from left to right in all figures) represent respectively pupil diameters of 5, 7.7, 10, 15 and \mbox{20 mm}. We see that high-contrast imaging within the dark hole at small separations (top plot) requires a large setup compared with larger dark holes (center and bottom plots). The figure also shows a strong dependence on the pupil diameter (large diameters - yellow and light blue curves - require a larger DM distance) compared with a weak dependence on the wavelength. When designing an optical high-contrast bench, this figure can be useful for a baseline solution for optics typical distances, because the overall setup length scales with the DM distances. 
The simulation in this paper is represented by the blue curve in the top figure, showing an optimum total distance of \mbox{2.7 m} at \mbox{1.65 $\mu$m}. This is consistent with the numerical results from end-to-end simulation as shown in Fig.~\ref{fig:discuss1} which represents the contrast ratio $\mathscr{C}$ as a function of the total distance between the two DMs for an aberration of 20 nm (black) and 40 nm (red). Solid lines correspond to the dispersion (number of random realizations that reach a contrast ratio greater that \mbox{5 $\sigma$} of the contrast values). The optimum DM distance (that reaches the best contrast) is about \mbox{3 m} and does not depend on the amount of aberration. We also see a significant contrast ratio degradation when increasing the aberration from 20 to \mbox{40 nm} (contrast ratio from $\thicksim10^{-14}$ to $10^{-12}$). Regarding the actual achievable contrast, the best contrast value should be computed for each specific set of parameters (according to the optimal DM distance presented in Fig.\ref{fig:discuss}), as there is no simple analytical way to assess the achievable contrast.
\\ The simulation in Section \ref{sec:model} and \ref{sec:results} assumes a perfect coronagraph and AO system, leading to resulting contrast values well below what can be achieved in realistic conditions. We nevertheless show that an inappropriate optical configuration (not optimum DM distances or large amount of high-frequency errors) can limit the results at the same order of magnitude as real AO systems or coronagraphs. In order to put these results into the context of future large telescope extreme performances, we simulate a pseudo-corrected wavefront corresponding to an AO system on a 40-m telescope diameter (Kolmogorov model) with \mbox{200 $\times$ 200} actuators (as expected for the E-ELT, \citealt{Kasper2012}). A more realistic coronagraph is not simulated, as we do not take into account special coronagraphic mask manufacturing defects, because they are specific to each coronagraph. The uncorrected high-frequency aberrations (beyond the DM cutoff frequency) severely degrade the achieved contrast to a level of $10^{-5}$, largely dominated by aliased speckles, consistent with Section \ref{sec:alias}. Although it is beyond the scope of this paper, we find that these high frequencies can be removed by placing a spatial filter (a simple hole) at the focal plane after the AO correction level but before the coronagraphic instrument. A spatial filter does not totally remove the high frequencies, because the diffraction effect of the filter itself creates high frequencies, but at a much lower level. This is the same principle as for the spatially filtered wavefront sensor described in \citet{PoyneerMacintosh2004} to reduce wavefront estimation errors resulting from aliasing. In our case, the simulation shows no performance impact when reducing the high-frequency errors by a factor of 10 or more. The next step could be to perform a full and more realistic simulation in the case of ground-based telescopes with atmospheric turbulence using this spatial filtering method. 

\section{Conclusion}
In this paper, we have assessed the main limitations to high-contrast imaging at small separations using wavefront control with two DMs. We first analysed, analytically or semi-analytically, some limitations to high-contrast imaging owing to DM location, actuator number and aberration magnitude and power spectrum (\textit{aliased} speckles). This analysis showed that high-contrast imaging around \mbox{1 $\lambda$/D} requires large inter-DM distances and small dark hole sizes. An in-depth simulation was developed to validate these theoretical results. The simulated model is based on a generic high-contrast test-bed combining a coronagraph and a wavefront control system with two DMs. The simulation takes into account the Fresnel propagation of static aberrations. We demonstrated that (1) the optimum DM location can be estimated analytically, (2) the dark hole algorithm is sensitive to the amount and the spatial distribution of optic aberrations (owing to aliased speckles and linear assumption of the algorithm), and (3) decreasing the dark hole size minimizes the performance dependence on actuator number and thus on the setup architecture (DM in convergent or collimated beams). We also demonstrated that high-contrast imaging at small IWA requires a small dark hole size to be consistent with the analytical approach of Section \ref{dmloc}. The setup optimization (DM location) depends on the required IWA with a strong difference in setup length when targeting high-contrast imaging at small or large separations. Althought this work is focused on high-contrast imaging at small separations, the analytical or semi-analytical approach can be used as a basis to define high-contrast imaging setup at any separation. Future laboratory test in Lagrange Laboratory will enable the validation of these results. 

\section*{Acknowledgements}
MB is supported by a joint grant from Observatoire de la C\^{o}te d'Azur and the R\'{e}gion PACA for her PhD fellowship. The authors are grateful to the computing center personal for their support and to the referee for useful comments.


\bibliographystyle{mnras}
\bibliography{test0}



\bsp	
\label{lastpage}
\end{document}